\documentclass[prd,preprint,tightenlines,superscriptaddress,floatfix,showpacs,preprintnumbers,nofootinbib,eqsecnum]{revtex4}

 \usepackage[dvips,final]{graphicx}
     \usepackage{epsfig}
      \usepackage{bm}

\def\Pom{{\bf I\!P}}

\begin{document}

\thispagestyle{empty} \preprint{\hbox{}} \vspace*{-10mm}

\title{Subdominant terms in the production of $c \bar c$ pairs \\ 
in proton-proton collisions}

\author{M.~{\L}uszczak}
\email{luszczak@univ.rzeszow.pl}
\affiliation{University of Rzesz\'ow, PL-35-959 Rzesz\'ow, Poland}

\author{R.~Maciu{\l}a}
\email{rafal.maciula@ifj.edu.pl}
\affiliation{Institute of Nuclear Physics PAN, PL-31-342 Cracow, Poland} 

\author{A.~Szczurek}
\email{antoni.szczurek@ifj.edu.pl}

\affiliation{Institute of Nuclear Physics PAN, PL-31-342 Cracow, Poland} 
\affiliation{University of Rzesz\'ow, PL-35-959 Rzesz\'ow, Poland}

\date{\today}

\begin{abstract}
At high-energies the gluon-gluon fusion is the dominant mechanism of 
$c \bar c$ production. This process was calculated in the NLO collinear 
as well as in the k$_t$-factorization approaches in the past.
We show that the present knowledge of gluon distributions
does not allow to make a precise predictions for $c \bar c$ 
production at LHC, in particular at forward rapidities.
In this paper we study production of $c \bar c$ pairs including several
subleading mechanisms. 
This includes:
$gg \to Q \bar Q$, $\gamma g \to Q \bar Q$, $g \gamma \to Q \bar Q$, 
$\gamma \gamma \to Q \bar Q$.
In this context we use MRST-QED 
parton distributions which include photon as a parton in the proton
as well as elastic photon distributions calculated in the
equivalent photon approximation.
We present distributions in the $c$ quark ($\bar c$ antiquark) rapidity 
and transverse momenta and compare them to the dominant gluon-gluon fusion
contribution.
We discuss also inclusive single and central diffractive processes using
diffractive parton distribution found from the analysis of HERA diffractive data.
As in the previous case we present distribution in $c$ ($\bar c$)
rapidity and transverse momentum.
Finally we present results for exclusive central diffractive mechanism
discussed recently in the literature. We show corresponding differential
distributions and compare them with corresponding distributions for
single and central diffractive components.
\end{abstract}


\maketitle

\section{Introduction}

In the past we have calculated inclusive cross section for 
heavy quarks production at hadron colliders. These calculations were 
performed using an approach based on the unintegrated parton distributions 
functions \cite{LS2006,luszczak2}. It is known that gluon-gluon fusion
is the dominant mechanism at high energy. However, other mechansims
were not carefully studied in the literature.

It is the aim of this work to present contributions of several
subleading terms usually neglected in the analysis of $c \bar c$
production. We wish to include contributions of photon-gluon
(gluon-photon) as well as purely electromagnetic contributions
of photon-photon fusion.

We wish to discuss also diffractive processes (single and central)
in the framework of Ingelman-Schlein model corrected for absorption.
Such a model was used in  estimation of several diffractive
processes \cite{diffractive_gauge_bosons,diffractive_dijets,
diffractive_open_charm,diffractive_Higgs,double_pomeron,double_diff}.

The absorption corrections turned out to be necessary to understand 
a huge Regge-factorization breaking observed in single and 
central production at Tevatron.

Recently a surprisingly large cross section for exclusive $c \bar c$
production has been reported \cite{MPS10}. Her we will show results for 
RHIC and LHC energies.

\section{Production of heavy quarks}

In the leading-order (LO) approximation within the collinear approach
the quadruply differential cross section in the rapidity 
of $Q$ ($y_1$),
in the rapidity of $\bar Q$ ($y_2$) and the transverse momentum of
one of them ($p_t$) can be written as
\begin{equation}
\frac{d \sigma}{d y_1 d y_2 d^2p_t} = \frac{1}{16 \pi^2 {\hat s}^2}
\sum_{i,j} x_1 p_i(x_1,\mu^2) \; x_2 p_j(x_2,\mu^2) \;
\overline{|{\cal M}_{ij \to Q \bar Q}|^2} \; .
\label{LO_collinear}
\end{equation}
Above, $p_i(x_1,\mu^2)$ and $p_j(x_2,\mu^2)$ are the familiar
(integrated) parton distributions in hadron $h_1$ and $h_2$, respectively.
There are two types of the LO $2 \to 2$ subprocesses which enter
Eq.(\ref{LO_collinear}): $gg \to Q \bar Q$ and $q \bar q \to Q \bar Q$.
The first mechanism dominates at large energies and the second one
near the threshold. 
In particular for the gluon-gluon fusion the cross section formula takes
a simple form:
\begin{equation}
\frac{d \sigma}{d y_1 d y_2 d^2p_t} = \frac{1}{16 \pi^2 {\hat s}^2}
x_1 g(x_1,\mu^2) \; x_2 g(x_2,\mu^2) \;
\overline{|{\cal M}_{gg \to Q \bar Q}|^2} \; .
\label{LO_collinear_gg}
\end{equation}
There are three ($s$, $t$ and $u$) diagrams in the leading order 
\cite{BP_book}.

\begin{figure*} %
\begin{center}
\includegraphics[width=4cm]{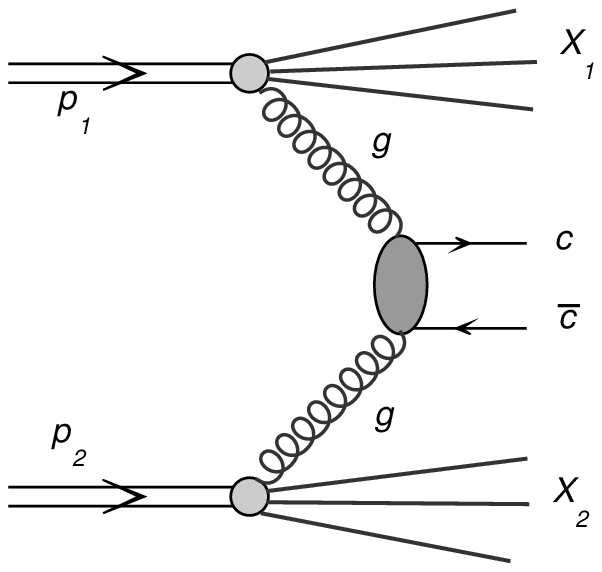}
\includegraphics[width=4cm]{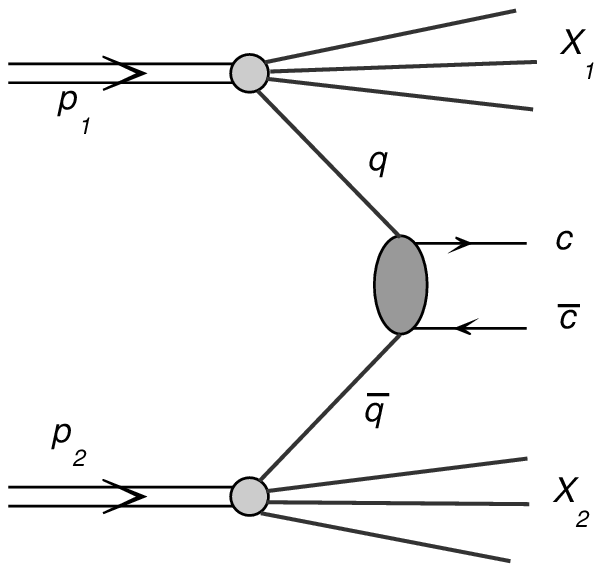}
\caption{Standard diagrams representing mechanisms 
for production of heavy quarks.
}
\label{fig:old_diagrams}
\end{center}
\end{figure*}


The parton distributions are evaluated at:
$x_1 = \frac{m_t}{\sqrt{s}}\left( \exp( y_1) + \exp( y_2) \right)$,
$x_2 = \frac{m_t}{\sqrt{s}}\left( \exp(-y_1) + \exp(-y_2) \right)$,
where $m_t = \sqrt{p_t^2 + m_Q^2}$.
The formulae for matrix element squared averaged over the initial
and summed over the final spin polarizations can be found e.g. in
Ref.\cite{BP_book}.

The inclusive heavy quark/antiquark production
can be also calculated in the framework of the $k_t$-factorization.
In this approach transverse momenta of initial partons are included 
and emission of gluons is encoded in so-called unintegrated 
gluon distributions (UGDFs) \cite{LS2006}.

In the leading-order (LO) approximation within the $k_t$-factorization approach
the quadruply differential cross section in the rapidity 
of $Q$ ($y_1$), in the rapidity of $\bar Q$ ($y_2$) and the transverse 
momentum of $Q$ ($p_1,t$) and $Q$ ($p_2,t$) can be written as
\begin{eqnarray}
\frac{d \sigma}{d y_1 d y_2 d^2p_{1,t} d^2p_{2,t}} =
\sum_{i,j} \; \int \frac{d^2 \kappa_{1,t}}{\pi} \frac{d^2 \kappa_{2,t}}{\pi}
\frac{1}{16 \pi^2 (x_1 x_2 s)^2} \; \overline{ | {\cal M}_{ij \to Q \bar Q} |^2}\\
\nonumber 
\delta^{2} \left( \vec{\kappa}_{1,t} + \vec{\kappa}_{2,t} 
                 - \vec{p}_{1,t} - \vec{p}_{2,t} \right) \;
{\cal F}_i(x_1,\kappa_{1,t}^2) \; {\cal F}_j(x_2,\kappa_{2,t}^2) \; , 
\nonumber  
\end{eqnarray}
where ${\cal F}_i(x_1,\kappa_{1,t}^2)$ and ${\cal F}_j(x_2,\kappa_{2,t}^2)$
are so-called unintegrated gluon (parton) distributions. 
Now the unintegrated parton distributions must be evaluated at:
$x_1 = \frac{m_{1,t}}{\sqrt{s}}\exp( y_1) 
     + \frac{m_{2,t}}{\sqrt{s}}\exp( y_2)$,
$x_2 = \frac{m_{1,t}}{\sqrt{s}}\exp(-y_1) 
     + \frac{m_{2,t}}{\sqrt{s}}\exp(-y_2)$,
where $m_{i,t} = \sqrt{p_{i,t}^2 + m_Q^2}$.

\section{Photon induced production of heavy quarks}

\subsection{MRST-QED parton distributions}

As discussed above the dominant contributions are initiated by gluons or
quarks and antiquarks. In general even photon can be a constituent
of the proton. This was considered only in one work by Martin, Roberts, 
Stirling and Thorne \cite{MRST04}. Below we repeat the main aspects of 
their formalism. 

The factorization of the QED-induced collinear divergences leads to 
QED-corrected evolution equations for the parton distributions of 
the proton \cite{MRST04}:
\begin{eqnarray}
{\partial q_i(x,\mu^2) \over \partial \log \mu^2} &=& {\alpha_S\over 2\pi}
\int_x^1 \frac{dy}{y} \Big\{
    P_{q q}(y)\; q_i(\frac{x}{y},\mu^2)
     +  P_{q g}(y)\; g(\frac{x}{y},\mu^2)\Big\}
\, \nonumber \\
&  + &
   {\alpha\over 2\pi} \int_x^1 \frac{dy}{y} \Big\{
    \tilde{P}_{q q}(y)\; e_i^2 q_i(\frac{x}{y},\mu^2)  +  P_{q \gamma}(y)\;
e_i^2 \gamma(\frac{x}{y},\mu^2)         \Big\}  \nonumber \\
{\partial g(x,\mu^2) \over \partial \log \mu^2} &=& {\alpha_S\over 2
\pi} \int_x^1 \frac{dy}{y} \Big\{
    P_{g q}(y)\; \sum_j q_j(\frac{x}{y},\mu^2) 
 + 
    P_{g g}(y)\; g(\frac{x}{y},\mu^2)\Big\}
\, \nonumber \\
   {\partial \gamma(x,\mu^2) \over \partial \log \mu^2}
& =   & {\alpha
\over 2\pi} \int_x^1 \frac{dy}{y} 
   \Big\{ P_{\gamma q}(y)\; \sum_j e_j^2\; q_j(\frac{x}{y},\mu^2) 
+ 
  P_{\gamma \gamma}(y)\; \gamma(\frac{x}{y},\mu^2) \Big\} \; ,
\end{eqnarray}
where
\begin{eqnarray}
{\tilde P}_{qq} = C_F^{-1} P_{qq}, & &   P_{\gamma q} = 
C_F^{-1} P_{g q}, \nonumber \\
P_{q\gamma} = T_R^{-1} P_{q g} , & &  P_{\gamma \gamma} = - 
\frac{2}{3}\; \sum_i e_i^2\; \delta(1-y) \nonumber
\end{eqnarray}
and the parton distributions fullfil momentum conservation:
\begin{equation}
  \int_0^1 dx\;  x\; \Big\{\sum_i q_i(x,\mu^2) + g(x,\mu^2) + \gamma(x,\mu^2)
     \Big\}  = 1 \; .
\end{equation}
%

\subsection{Mechanisms with one or two photons}

\begin{figure*}
\begin{center}
\includegraphics[width=3cm]{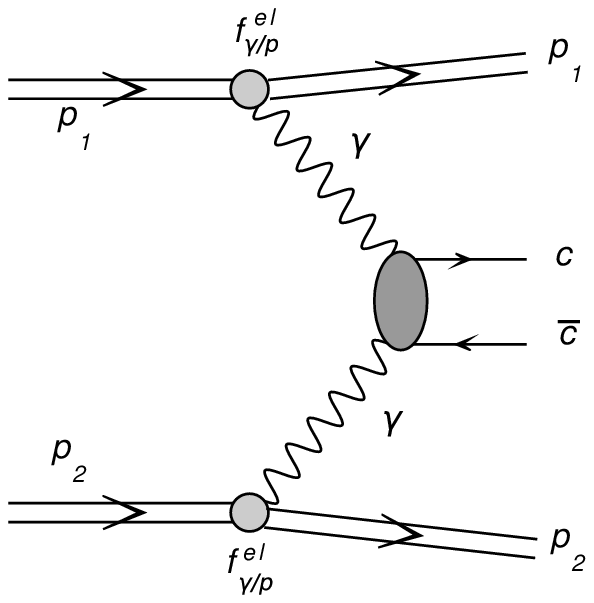}
\includegraphics[width=3cm]{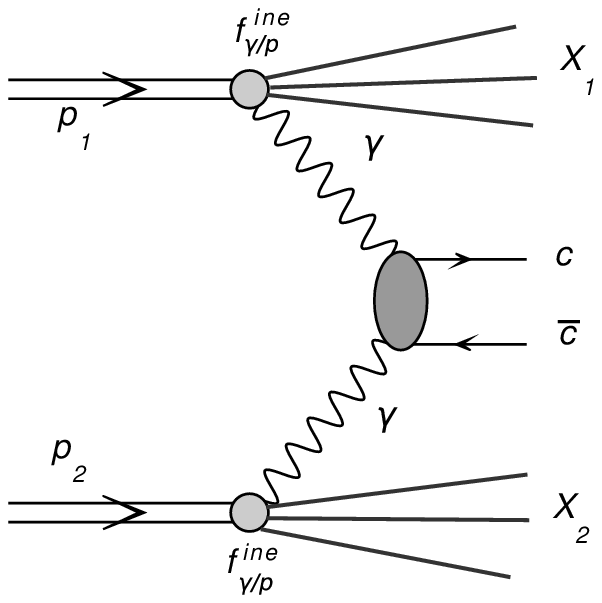}
\includegraphics[width=3cm]{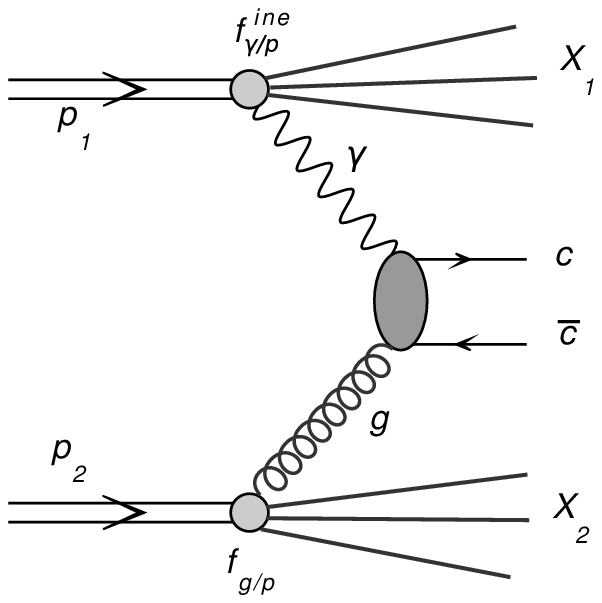}
\includegraphics[width=3cm]{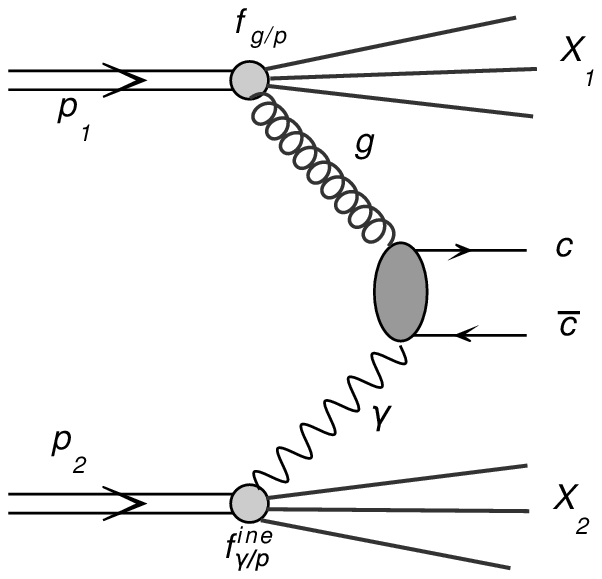}\\

\includegraphics[width=3cm]{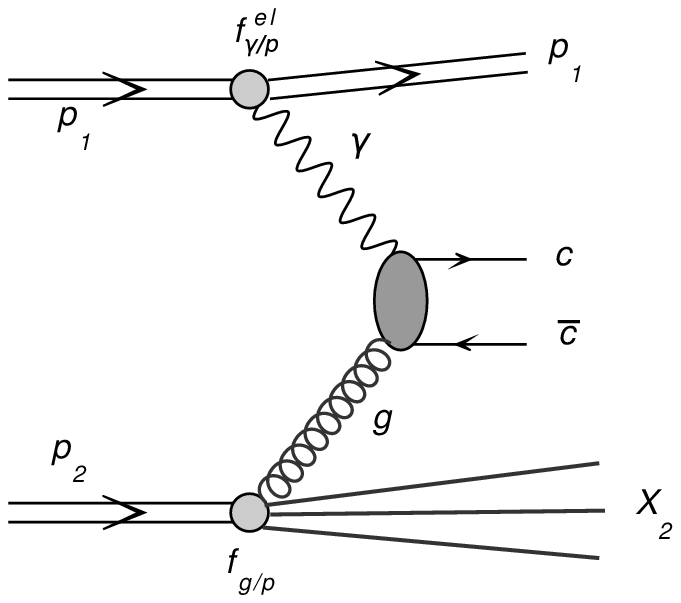}
\includegraphics[width=3cm]{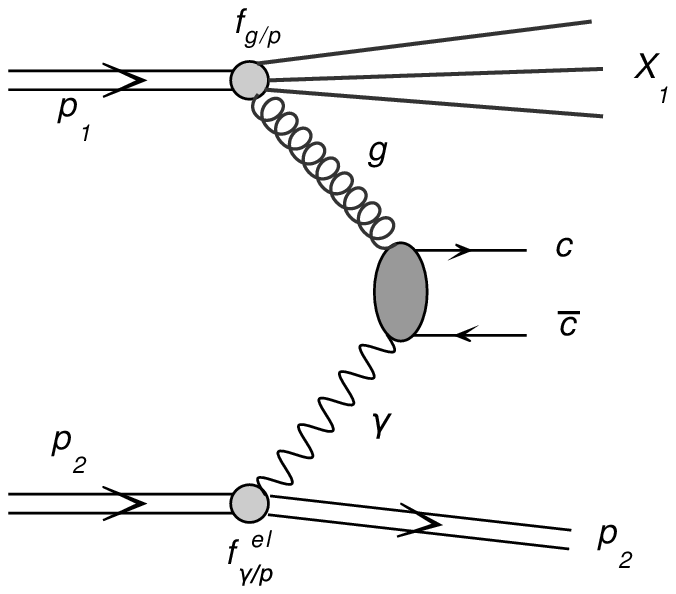}
\includegraphics[width=3cm]{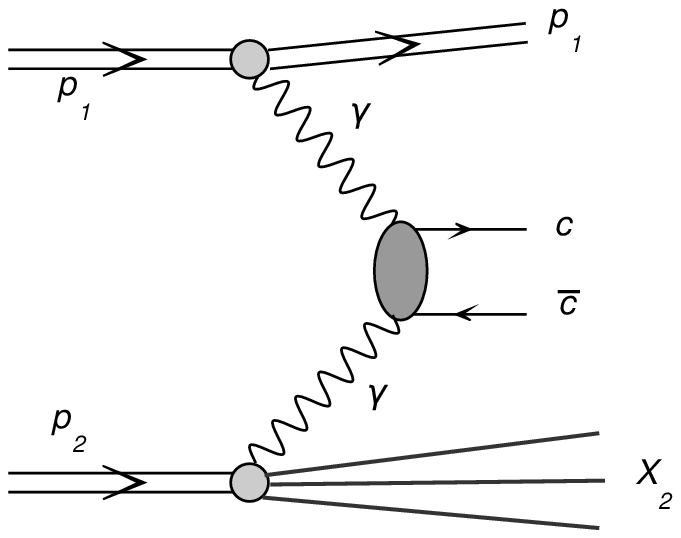}
\includegraphics[width=3cm]{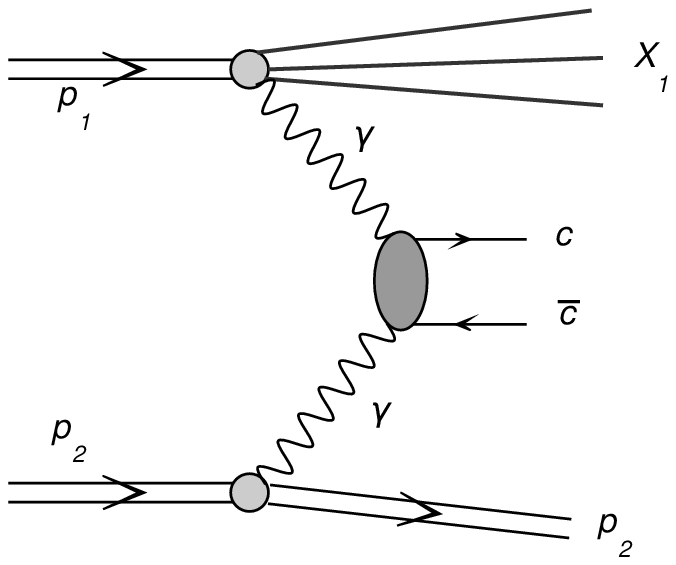}\\
\caption{Diagrams representing mechanisms 
for production of heavy quarks, which included photons.
}
\label{fig:new_diagrams}
\end{center}
\end{figure*}

If the photon is a constituent of the nucleon then other mechanisms
presented in Fig.\ref{fig:new_diagrams} are possible.

Here the cross section can be calculated similarly as for the
gluon-gluon fusion.
A corresponding triple differential cross section can be written as:
\begin{eqnarray}
\frac{d \sigma^{g\gamma_{in}}}{d y_1 d y_2 d^2p_t} &=& \frac{1}{16 \pi^2 {\hat s}^2}
x_1 g(x_1,\mu^2) \; x_2 \gamma_{in}(x_2,\mu^2) \;
\overline{|{\cal M}_{g\gamma \to Q \bar Q}|^2} \; , \nonumber \\
\frac{d \sigma^{\gamma_{in} g}}{d y_1 d y_2 d^2p_t} &=& \frac{1}{16 \pi^2 {\hat s}^2}
x_1 \gamma_{in}(x_1,\mu^2) \; x_2 g(x_2,\mu^2) \;
\overline{|{\cal M}_{\gamma g \to Q \bar Q}|^2} \; , \nonumber \\
\frac{d \sigma^{\gamma_{in} \gamma_{in}}}{d y_1 d y_2 d^2p_t} &=& \frac{1}{16 \pi^2 {\hat s}^2}
x_1 \gamma_{in}(x_1,\mu^2) \; x_2 \gamma_{in}(x_2,\mu^2) \;
\overline{|{\cal M}_{\gamma \gamma \to Q \bar Q}|^2} \;
\end{eqnarray}
for gluon-photon, photon-gluon and photon-photon contributions, respectively.
Compared to gluon-gluon case here only $t$ and $u$ diagrams occur.

The above contributions include only cases when nucleons do not survive
a collision and nucleon debris is produced instead. The case when
nucleon survives a collision has to be considered separately. In this
case one can include corresponding photon distributions where extra "el"
index will be added to denote that situation. Corresponding
contributions can be then written as:
\begin{eqnarray}
\frac{d \sigma^{g\gamma_{el}}}{d y_1 d y_2 d^2p_t} &=& \frac{1}{16 \pi^2 {\hat s}^2}
x_1 g(x_1,\mu^2) \; x_2 \gamma_{el}(x_2,\mu^2) \;
\overline{|{\cal M}_{g\gamma \to Q \bar Q}|^2} \; , \nonumber \\
\frac{d \sigma^{\gamma_{el} g}}{d y_1 d y_2 d^2p_t} &=& \frac{1}{16 \pi^2 {\hat s}^2}
x_1 \gamma_{el}(x_1,\mu^2) \; x_2 g(x_2,\mu^2) \;
\overline{|{\cal M}_{\gamma g \to Q \bar Q}|^2} \; , \nonumber \\
\frac{d \sigma^{\gamma_{in} \gamma_{el}}}{d y_1 d y_2 d^2p_t} &=& \frac{1}{16 \pi^2 {\hat s}^2}
x_1 \gamma_{in}(x_1,\mu^2) \; x_2 \gamma_{el}(x_2,\mu^2) \;
\overline{|{\cal M}_{\gamma \gamma \to Q \bar Q}|^2} \; ,\nonumber \\
\frac{d \sigma^{\gamma_{el} \gamma_{in}}}{d y_1 d y_2 d^2p_t} &=& \frac{1}{16 \pi^2 {\hat s}^2}
x_1 \gamma_{el}(x_1,\mu^2) \; x_2 \gamma_{in}(x_2,\mu^2) \;
\overline{|{\cal M}_{\gamma \gamma \to Q \bar Q}|^2} \; ,\nonumber \\
\frac{d \sigma^{\gamma_{el} \gamma_{el}}}{d y_1 d y_2 d^2p_t} &=& \frac{1}{16 \pi^2 {\hat s}^2}
x_1 \gamma_{el}(x_1,\mu^2) \; x_2 \gamma_{el}(x_2,\mu^2) \; 
\overline{|{\cal M}_{\gamma \gamma \to Q \bar Q}|^2} \; . \\ 
\label{subleading_contributions}
\end{eqnarray}
The elastic contributions are calculated using Drees-Zepenfeld 
(elastic) parametrizations of photon fluxes \cite{DZ1989} which include 
nucleon electromagnetic form factors.

\section{Results}

\subsection{Gluon distributions and small-x region and
its relation to $c \bar c$ production}

\begin{figure*} [!thb]
\includegraphics[width=6cm]{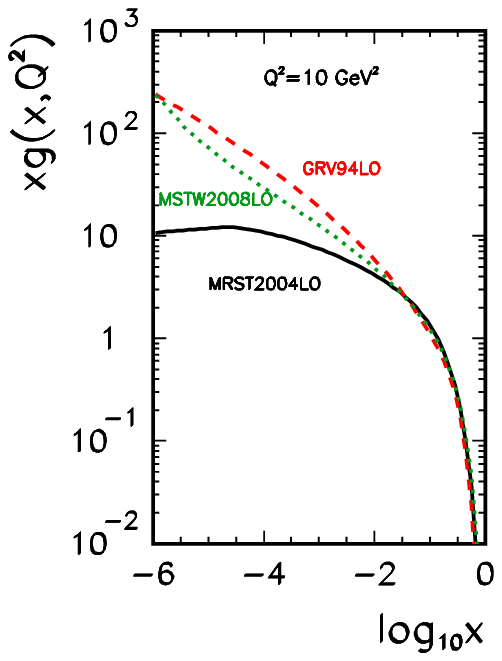}
\includegraphics[width=6cm]{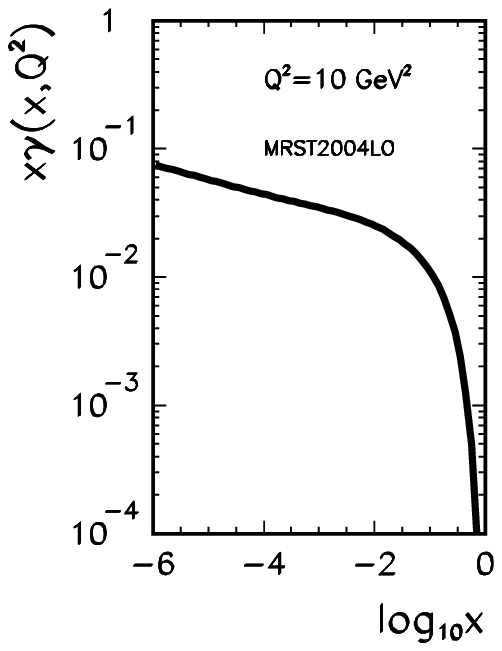}
\caption{\small Different leading order gluon distributions
form the literature for the factorization scale:
$\mu^2$ = 10 GeV$^2$ (left panel) and leading order photon distributions
for factorization scale: $\mu^2$ = 10 GeV$^2$ (right panel).}
\label{fig:xg}
\end{figure*}

In Fig.\ref{fig:xg} we show three different leading-order gluon 
distributions from the literature \cite{GRV94,MRST04,MSTW08} (left
panel) and photon distributions \cite{MRST04} (right panel)
as a function of longitudinal momentum fraction $x$ for a fixed scale
$\mu^2 =$ 10 GeV relevant for $c \bar c$ production. Above 
$x >$ 10$^{-2}$ all the distributions coincide. For smaller values of 
$x$ they diverge and can be different by almost an order of magnitude.
What are consequences of this divergence for $c \bar c$ pair production?
This will be discussed below.

\begin{figure*} [!thb]
\begin{center}
\includegraphics[width=6cm]{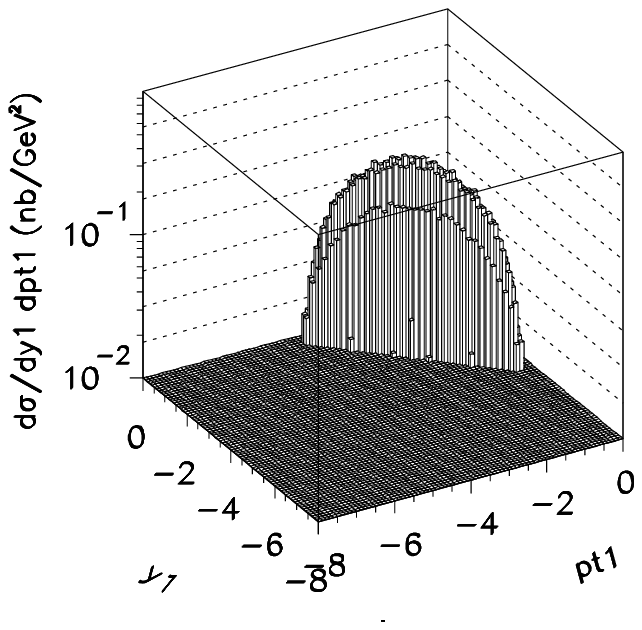}
\includegraphics[width=6cm]{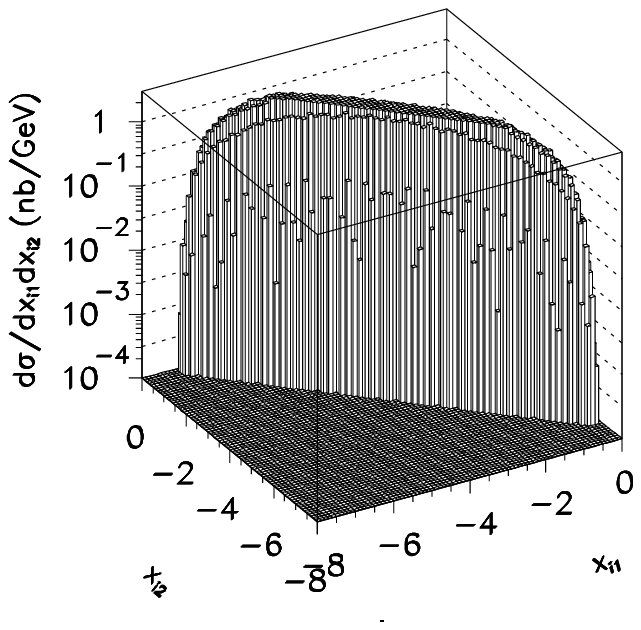}
\end{center}
\caption{\small Distributions in $x_1$ and $x_2$
for two different energies: $\sqrt{s}$ = 500 GeV (left)
and $\sqrt{s}$ = 14000 GeV (right).
In this calculation GRV94 gluon distributions have been used.
}
\label{fig:dsig_dxi1dxi2}
\end{figure*}

Before we go to cross sections in transverse momentum and rapidities, 
in Fig.\ref{fig:dsig_dxi1dxi2} we present distribution of the cross
section in $\xi_1$ = log$_{10} x_1$ and $\xi_2$ = log$_{10} x_2$ for
two different energies $\sqrt{s}$ = 500 GeV (updated RHIC)
and $\sqrt{s}$ = 14 TeV (nominal LHC energy).
One can clearly see that the $x_1$ and $x_2$ values are strongly
correlated. Typical values at $\sqrt{s}$ = 500 GeV are $x_1, x_2 \sim$
0.5 $\cdot$ 10$^{-2}$ and at $\sqrt{s}$ = 14 TeV are $x_1, x_2 \sim$
10$^{-4}$. In the latter case $x$'s as small as 10$^{-6}$ may appear
in the forward $c$ or $\bar c$ region.
This is clearly a region of $x$ which was never studied so far.

\begin{figure*} [!thb]
\begin{center}
\includegraphics[width=5cm]{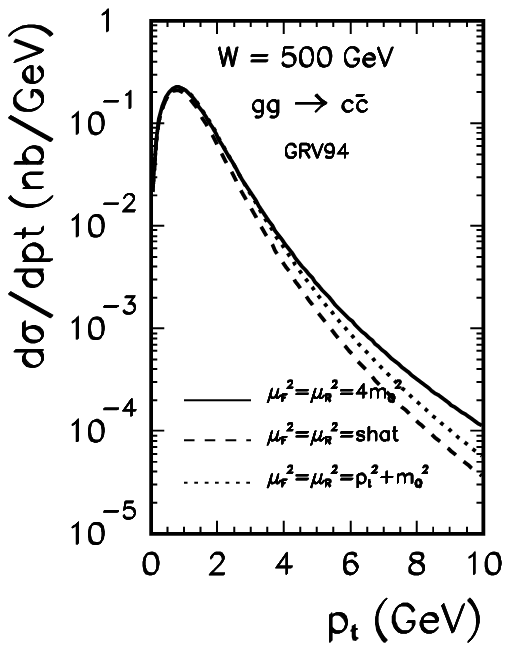}
\includegraphics[width=5cm]{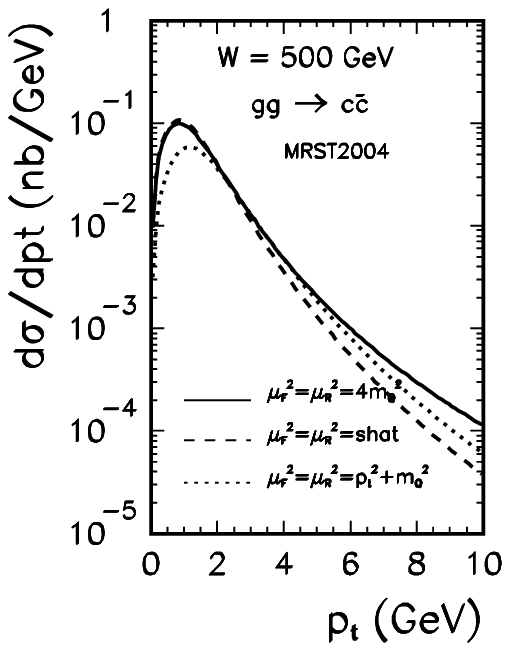}
\includegraphics[width=5cm]{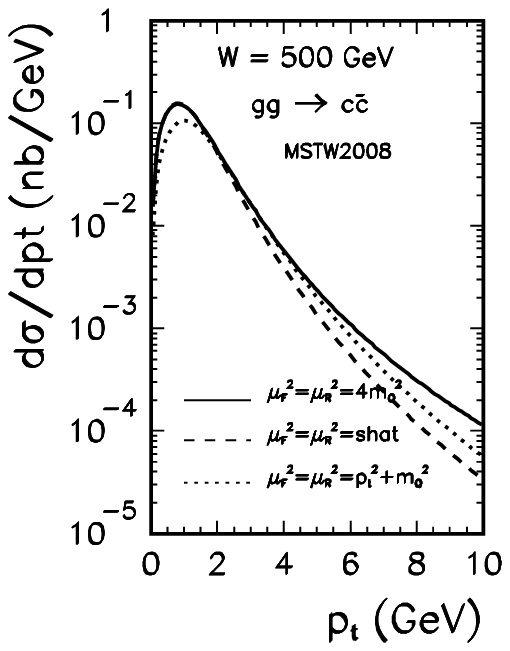}
\end{center}
\caption{\small Distribution in quark/antiquark 
transverse momentum at $\sqrt{s}$ = 500 GeV for different choices of 
scales and for different gluon distributions:
GRV94 (left panel), MRST2004 (midle panel) and MSTW2008 (right panel).
In this calculation we have used $\mu_F^2 = \mu_R^2 = 4 m_Q^2$.
}
\label{fig:dsig_dpt_500}
\end{figure*}

\begin{figure*} [!thb]
\begin{center}
\includegraphics[width=5cm]{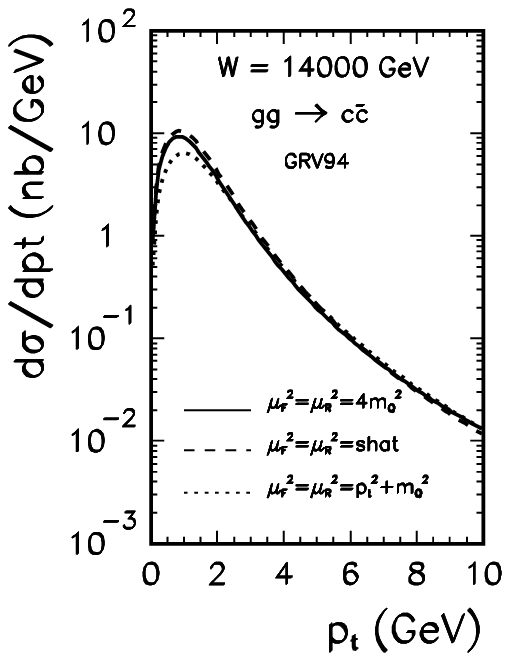}
\includegraphics[width=5cm]{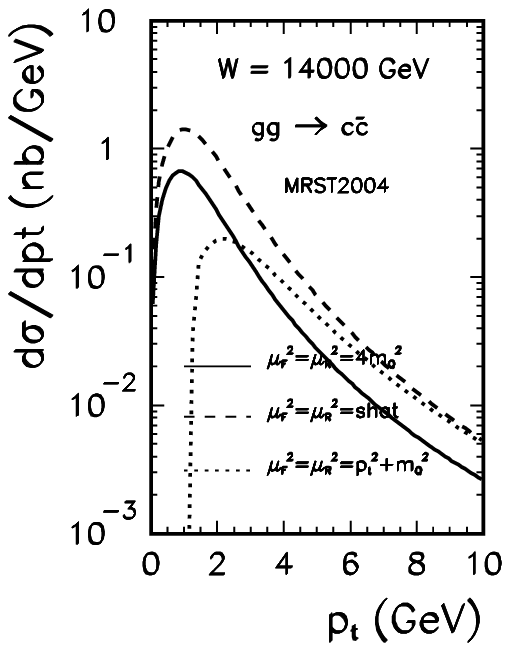}
\includegraphics[width=5cm]{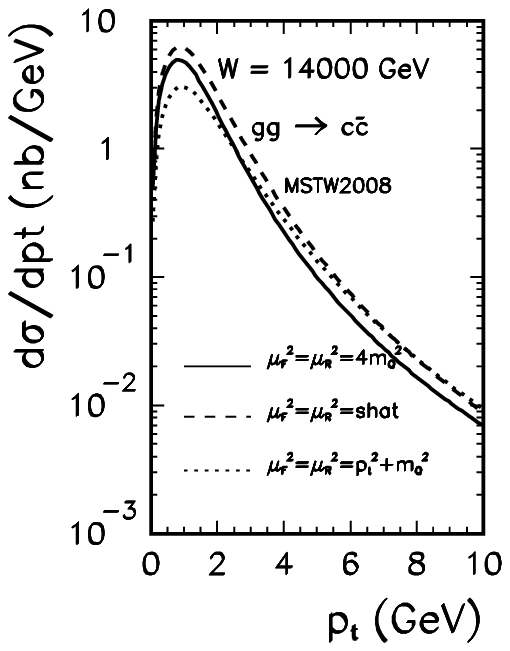}
\end{center}
\caption{Distribution in quark/antiquark 
transverse momentum at $\sqrt{s}$ = 14 TeV for different choices 
of scales and for different gluon distributions:
GRV94 (left panel), MRST2004 (midle panel) and MSTW2008 (right panel).
In this calculation we use have used $\mu_F^2 = \mu_R^2 = 4 m_Q^2$.
}
\label{fig:dsig_dpt_14000}
\end{figure*}

Now let us present distributions in transverse momentum of 
$c$ (or $\bar c$) for gluon-gluon fusion mechanism
for different gluon distributions and different popular choices of 
scales ($\mu^2 = 4 m_c^2$, invariant mass of the $c \bar c$ system  
$M_{c \bar c}^2, p_t^2 + m_c^2$). We show our results for $\sqrt{s}$ = 500 GeV
(Fig.\ref{fig:dsig_dpt_500}) and $\sqrt{s}$ = 14 TeV 
(Fig.\ref{fig:dsig_dpt_14000}).
One can clearly see that for some choices of gluon distribution function
and scales the results for $\sqrt{s}$ = 14 TeV are not physical. 
This shows how badly known are gluon distributions at the low $x$.

\subsection{$\gamma g$ and $g \gamma$ subprocesses}

\begin{figure*} [!thb]
\begin{center}
\includegraphics[width=6cm]{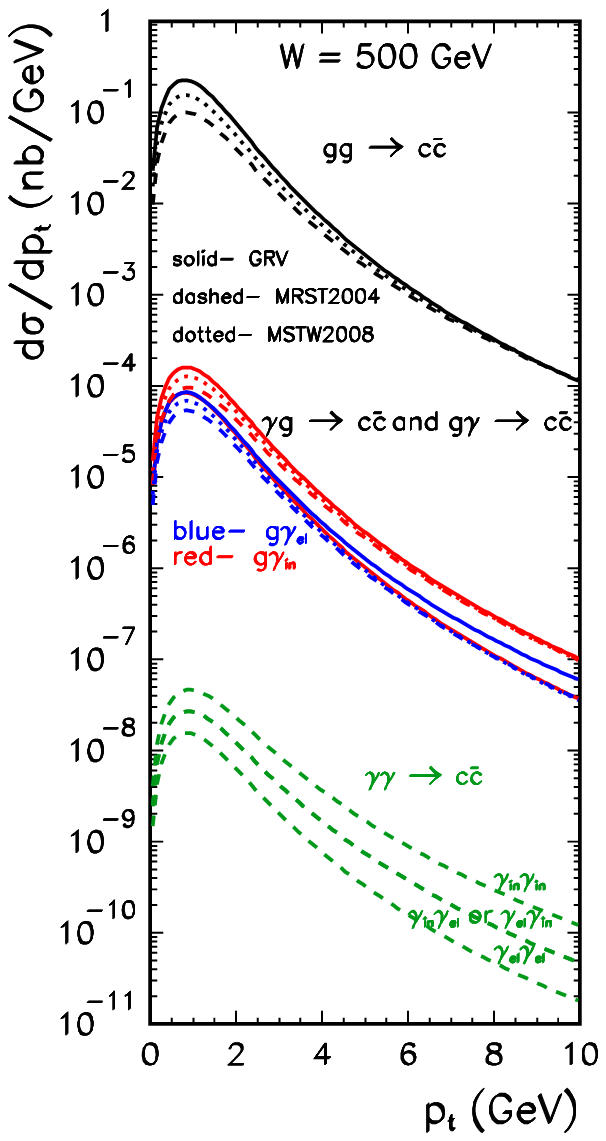}
\includegraphics[width=6cm]{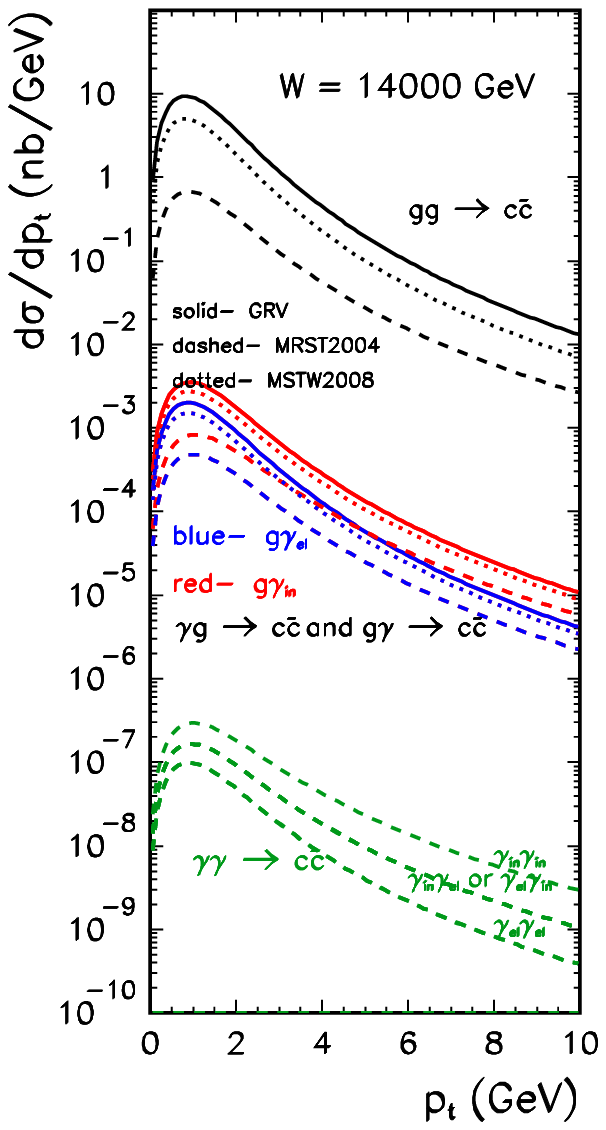}
\end{center}
\caption{
Transverse momentum distribution for the standard gluon-gluon
mixed gluon-photon and photon-gluon as well as for photon-photon
contributions for RHIC (left panel) and LHC (right panel). 
Three different gluon distributions were used. 
The photon distributions are from \cite{MRST04}.
We show contributions when proton survives the collision (called elastic)
and when hadronic debris is produced (called inelastic).
\label{fig:dsig_dpt_gamma}
}
\end{figure*}

\begin{figure*} [!thb]
\begin{center}
\includegraphics[width=5cm]{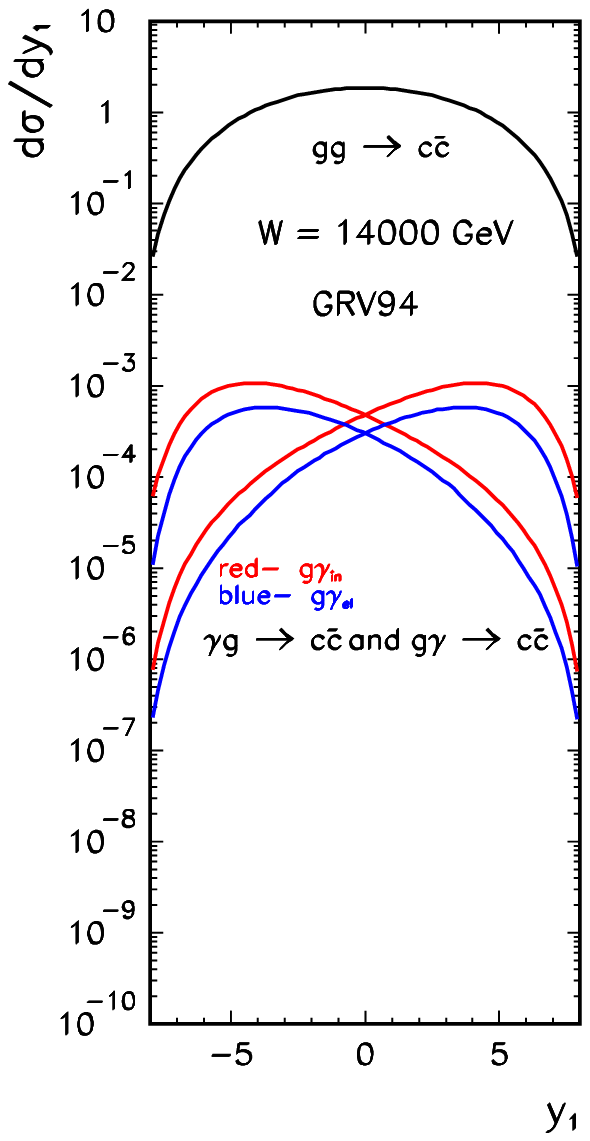}
\includegraphics[width=5cm]{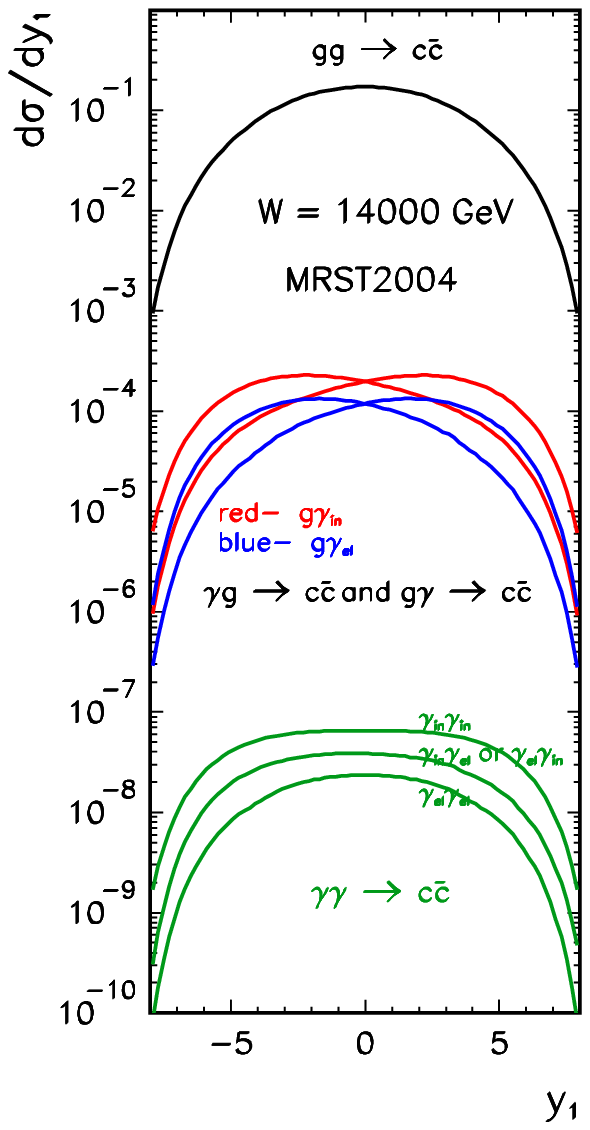}
\includegraphics[width=5cm]{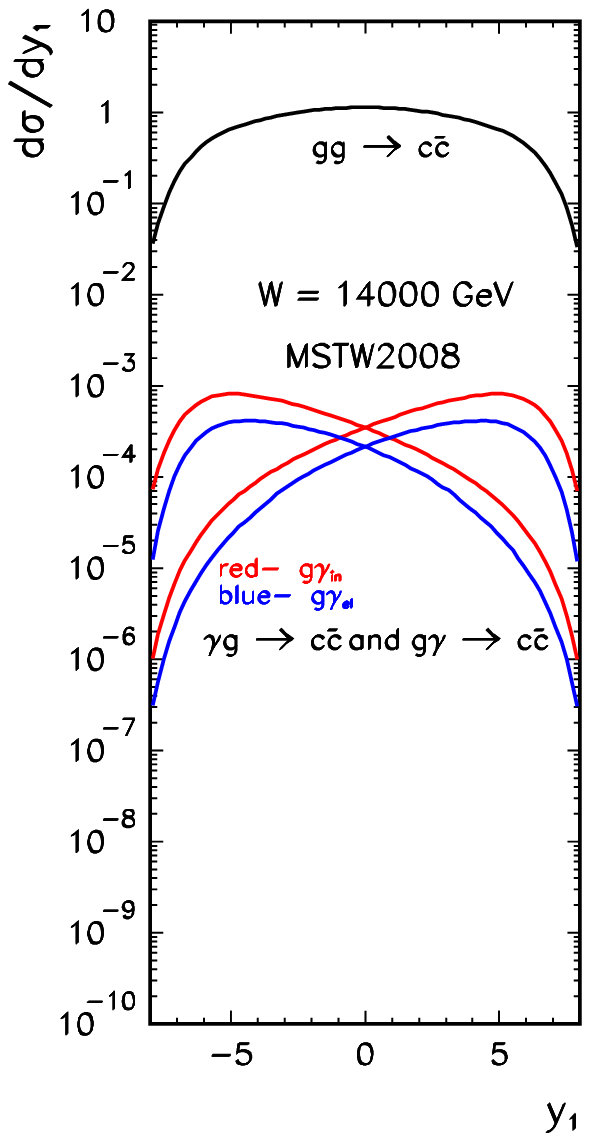}
\end{center}
\caption{Different contributions to distributions in rapidity 
of $c$ quark/antiquark at $\sqrt{s}$ = 14 TeV 
for different gluon distributions:
GRV94 (left panel), MRST2004 (midle panel) and MSTW2008 (right panel).
In this calculation we have used $\mu_F^2 = \mu_R^2 = \hat{s}$.
\label{fig:dsig_dy_gamma}
}
\end{figure*}

In Fig.\ref{fig:dsig_dpt_gamma} and
in Fig.\ref{fig:dsig_dy_gamma},
we show results for different gluon distribution functions for
the RHIC energy $\sqrt{s}$ = 500 GeV and nominal LHC energy 
$\sqrt{s}$ = 14 TeV, respectively.
At the LHC energy the results for different GDFs differ considerably which
is a consequence of the small-$x$ region as discussed in the previous section.
The differences at the nominal LHC energy $\sqrt{s}$ = 14 TeV are 
particularly large which can be explained by the fact that a product of
gluon distributions (both at small $x$) enters the cross section formula.
A new measurement of $c \bar c$ at the nominal LHC energy will be
therefore a severe test of gluon distributions at small $x$ and not too
high factorization scales not tested so far. Similar uncertainties for 
the $\gamma g$ and $g \gamma$ are smaller as here only one gluon distribution
appears in the corresponding cross section formula.

It is very difficult to quantify uncertainties related to the photon PDFs
as only one set of PDFs includes photon as a parton of the proton.
Here the isospin symmetry violation (not well known at present)
would be an useful limitation.
Our collection of the results for the photon induced mechanisms show
that they are rather small and their identification would be rather
difficult as the different distributions are very similar to those
for the gluon-gluon fusion.
Our intension here is to document all the subleading terms in one
publication.
Our etimation shows that the sum of all the photon induced terms
is less than 0.5 \% and is by almost 2 orders of magnitude smaller
than the uncertainties of the dominant leading-order
gluon-gluon term.

\section{Single and central diffraction}

\label{sec:diffractive_production}

\subsection{Formalism}

\begin{figure}[!h]    %
\includegraphics[width=0.35\textwidth]{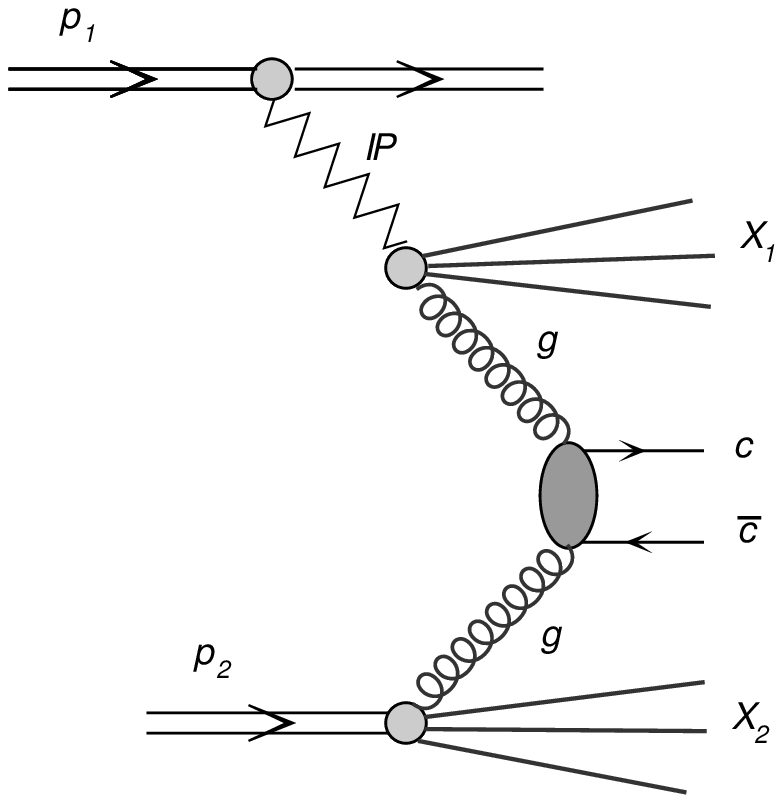}
\hspace{0.5cm}
\includegraphics[width=0.35\textwidth]{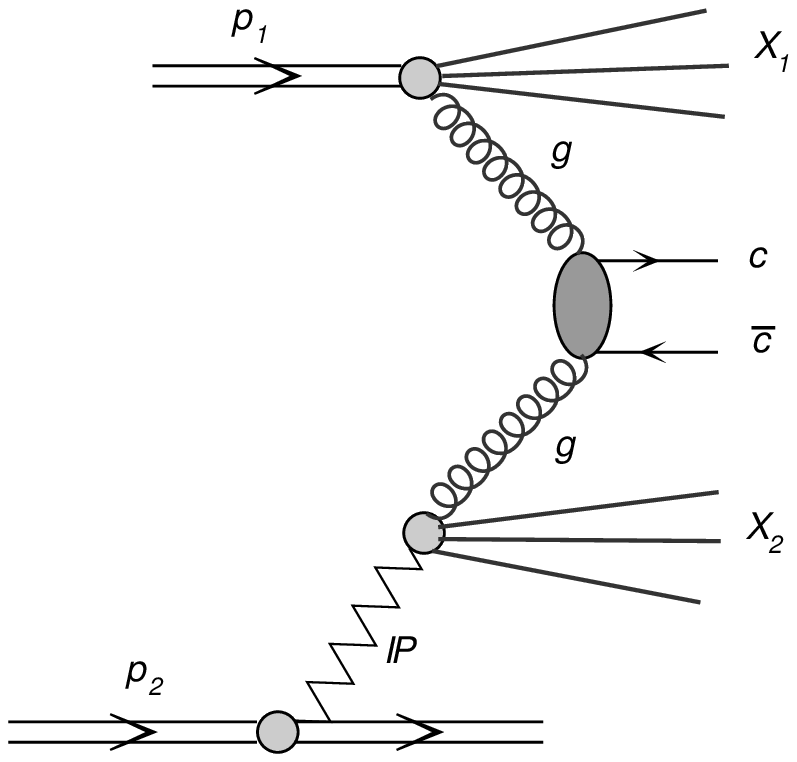}
   \caption{\label{fig:mechanism_sd}
   \small The mechanism of single-diffractive production of $c \bar c$.  
}
\end{figure}

\begin{figure}[!h]    %
\includegraphics[width=0.35\textwidth]{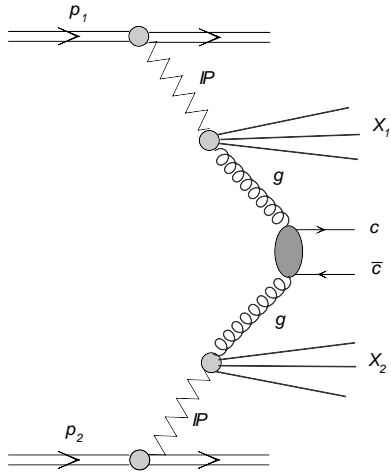}
   \caption{\label{fig:mechanism_dd}
   \small The mechanism of central-diffractive production of dileptons.  
}
\end{figure}

The mechanisms of the ordinary as well as diffractive production 
of heavy quarks ($c \bar c$) are shown in 
Figs.\ref{fig:mechanism_sd},\ref{fig:mechanism_dd}.

In the following we apply the Ingelman and Schlein approach\footnote{In the literature
also dipole model was used to estimate diffractive $c \bar c$ production 
\cite{Kopeliovich}. }.
In this approach one assumes that the Pomeron has a
well defined partonic structure, and that the hard process
takes place in a Pomeron--proton or proton--Pomeron (single diffraction) 
or Pomeron--Pomeron (central diffraction) processes.
We calculate triple differential distributions as
\begin{eqnarray}
{d \sigma_{00} \over dy_{1} dy_{2} dp_{t}^2} =  K {\Big| M \Big|^2 \over 16 \pi^2 \hat{s}^2} 
\,\Big [\, \Big( x_1 q_f(x_1,\mu^2) 
\, x_2 \bar q_f(x_2,\mu^2) \Big) \, 
+ \Big( x_1 \bar q_f(x_1,\mu^2)
\, x_2  q_f(x_2,\mu^2) \Big) \, \Big ] ,
\nonumber \\ 
\label{00}
\end{eqnarray}
\begin{eqnarray}
{d \sigma_{SD} \over dy_{1} dy_{2} dp_{t}^2} =  K {\Big| M \Big|^2 \over 16 \pi^2 \hat{s}^2} 
\,\Big [\, \Big( x_1 q_f^D(x_1,\mu^2) 
\, x_2 \bar q_f(x_2,\mu^2) \Big) \, 
+ \Big( x_1 \bar q_f^D(x_1,\mu^2)
\, x_2  q_f(x_2,\mu^2) \Big) \, \Big ] ,
\nonumber \\ 
\label{SD}
\end{eqnarray}
\begin{eqnarray}
{d \sigma_{CD} \over dy_{1} dy_{2} dp_{t}^2} =  K {\Big| M \Big|^2 \over 16 \pi^2 \hat{s}^2} 
\,\Big [\, \Big( x_1 q_f^D(x_1,\mu^2) 
\, x_2 \bar q_f^D(x_2,\mu^2) \Big) \, 
+ \Big( x_1 \bar q_f^D(x_1,\mu^2)
\, x_2  q_f^D(x_2,\mu^2) \Big) \,\Big ] 
\nonumber \\ 
\label{DD}
\end{eqnarray}
for ordinary, single-diffractive and central-diffractive production, 
respectively.

We do not calculate the higher-order contributions and include them 
effectively with the help of a so-called $K$-factor. We have checked
that this procedure is precise enough in the case of ordinary Drell-Yan
process.
The $K$-factor is calculated as for the Drell-Yan process
\begin{eqnarray*}
K &=& 1 +{ \alpha_{s}  \over  2 \pi}{ 4 \over 3} \Big (1+ { 4 \over 3} \pi^2 \Big).
\end{eqnarray*}
Here the running coupling constant $\alpha_{s} = \alpha_{s}(\mu^2)$ is
evaluated at $\mu^2 = M_{Q \bar Q}^2$.

The 'diffractive' quark distribution of
flavour $f$ can be obtained by a convolution of the flux of Pomerons
$f_\Pom(x_\Pom)$ and the parton distribution in the Pomeron 
$q_{f/\Pom}(\beta, \mu^2)$:
\begin{eqnarray}
q_f^D(x,\mu^2) = \int d x_\Pom d\beta \, \delta(x-x_\Pom \beta) 
q_{f/\Pom} (\beta,\mu^2) \, f_\Pom(x_\Pom) \, 
= \int_x^1 {d x_\Pom \over x_\Pom} \, f_\Pom(x_\Pom)  
q_{f/\Pom}({x \over x_\Pom}, \mu^2) \, . \nonumber \\
\end{eqnarray}
The flux of Pomerons $f_\Pom(x_\Pom)$ enters in the form integrated over 
four--momentum transfer 
\begin{eqnarray}
f_\Pom(x_\Pom) = \int_{t_{min}}^{t_{max}} dt \, f(x_\Pom,t) \, ,
\label{flux_of_Pom}
\end{eqnarray}
with $t_{min}, t_{max}$ being kinematic boundaries.

Both pomeron flux factors $f_{\Pom}(x_{\Pom},t)$ as well 
as quark/antiquark distributions in the pomeron were taken from 
the H1 collaboration analysis of diffractive structure function
and diffractive dijets at HERA \cite{H1}. 
The factorization scale for diffractive parton distributions is taken as
$\mu^2 = \hat s$.

\subsection{Results}

Let us start presentation of our results for diffractive mechanisms.

\begin{figure*} [!thb]
\begin{center}
\includegraphics[width=5cm]{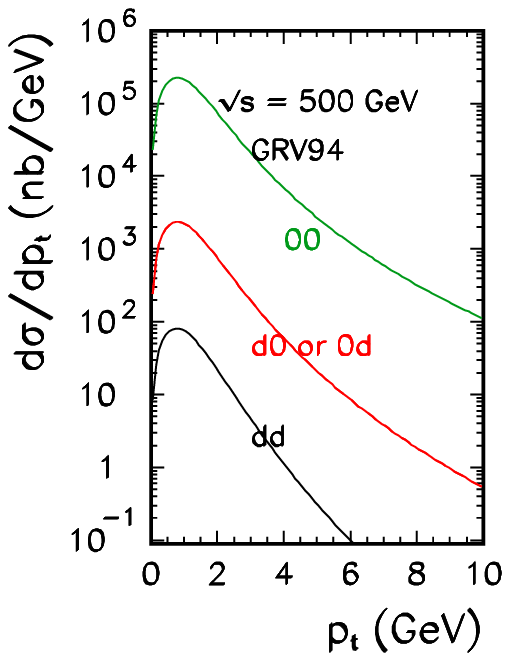}
\includegraphics[width=5cm]{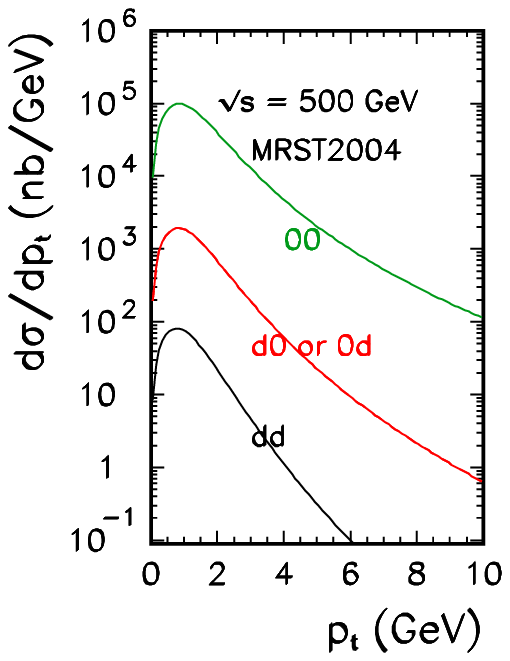}
\includegraphics[width=5cm]{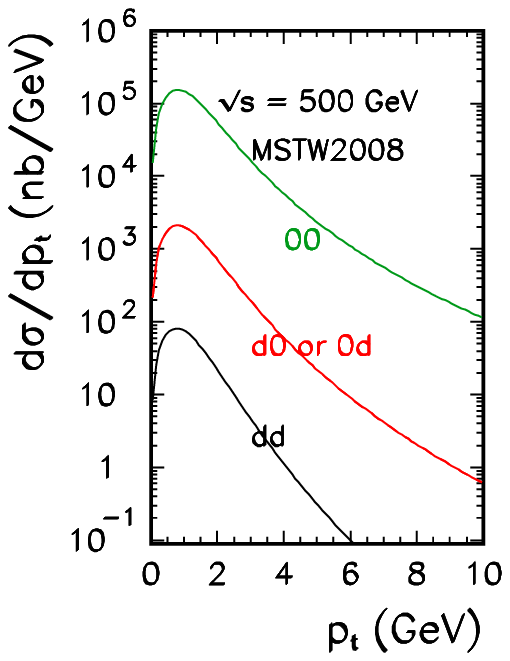}
\end{center}
\caption{\small Transverse momentum distribution of $c$ quarks (antiquarks)
for RHIC energy $\sqrt{s} =$ 500 GeV for three different parton
distributions. 
The result for single diffractive (0d or d0), central diffractive (dd) 
mechanisms are compared with the standard gluon-gluon fusion contribution (00).
\label{fig:diff_dsig_dpt_500}
}
\end{figure*}

In Fig.\ref{fig:diff_dsig_dpt_500} we show transverse momentum
distributions of charm quarks (or antiquarks). The distribution
for single diffractive component is smaller than that for the
inclusive gluon-gluon fusion by almost two orders of magnitude. Our
results include gap survival factor. Corresponding values are taken the
same as in Ref. \cite{diffractive_leptons}. The cross section for
inclusive central diffractive component is smaller by additional order 
of magnitude. In addition we show the cross section for fully\footnote{Although the
 calculation assumes simple $c \bar c$ state hadronization leads to more
complicated states \cite{MPS10}.}
exclusive mechanism discussed in section \ref{sec:exclusive_production}.
Below we shall use the following notation: 
$00$ for standard nondiffractive component,
$0d$ or $d0$ for single diffractive 
and $dd$ for central diffractive components.


\begin{figure*} [!thb]
\begin{center}
\includegraphics[width=5cm]{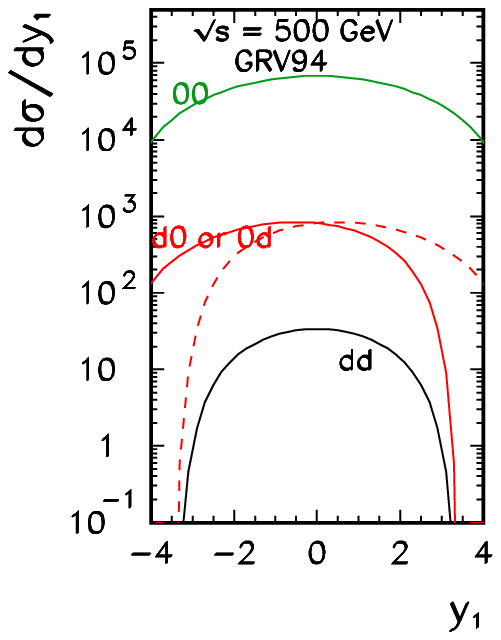}
\includegraphics[width=5cm]{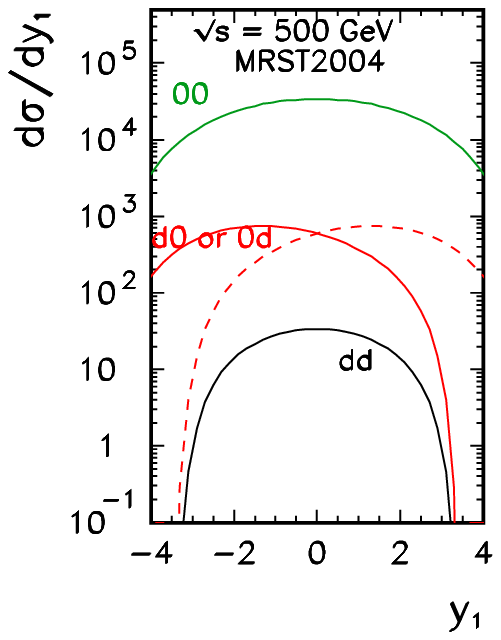}
\includegraphics[width=5cm]{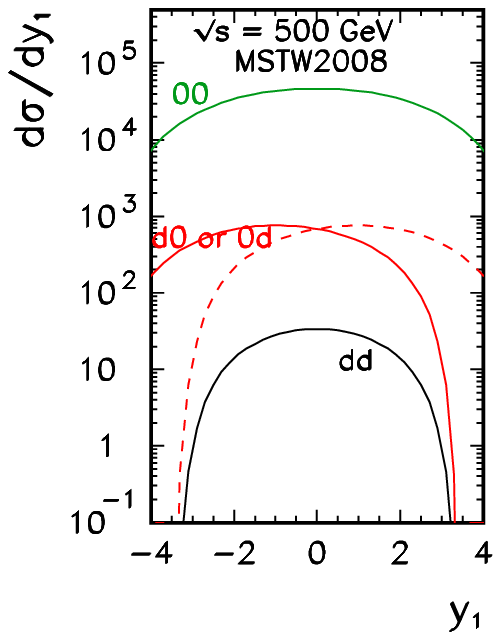}
\end{center}
\caption{\small
Rapidity distribution of $c$ quarks (antiquarks) for RHIC energy 
$\sqrt{s} =$ 500 GeV for three different parton distributions. 
The result for single diffractive (0d or d0), central diffractive (dd) 
mechanisms are compared with the standard gluon-gluon fusion contribution (00).
\label{fig:diff_dsig_dy1_500}
}
\end{figure*}

\begin{figure*} [!thb]
\begin{center}
\includegraphics[width=5cm]{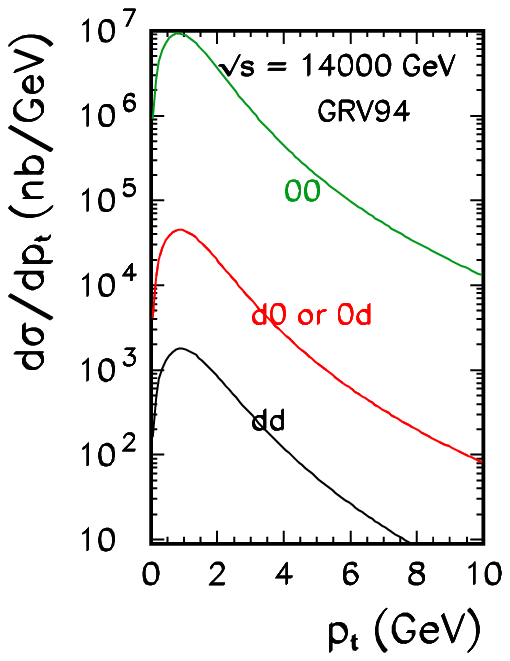}
\includegraphics[width=5cm]{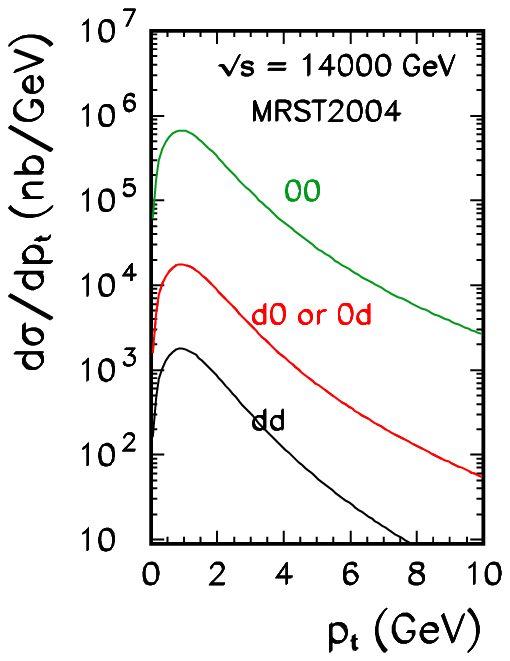}
\includegraphics[width=5cm]{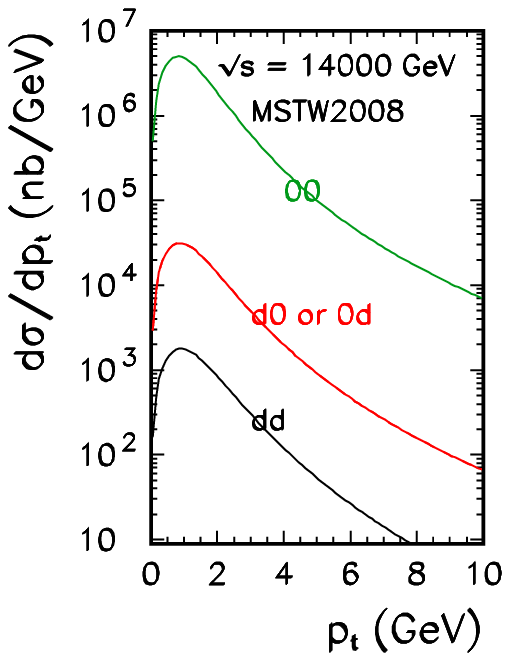}
\end{center}
\caption{\small
Transverse momentum distribution of $c$ quarks (antiquarks)
for RHIC energy $\sqrt{s} =$ 14000 GeV for three different parton
distributions 
The result for single diffractive (0d or d0), 
central diffractive (dd) mechanisms
are compared with the standard gluon-gluon fusion contribution (00).
\label{fig:diff_dsig_dpt_14000}
}
\end{figure*}

In Fig.\ref{fig:diff_dsig_dpt_14000} we show similar results for
nominal LHC energy $\sqrt{s}$ = 14 TeV. The situation and the
interrelations between different components is qualitatively the same.
Here somewhat smaller gap survival factors were used 
\cite{diffractive_leptons}. The distributions for all components
are somewhat broader than those for the RHIC energy shown above.


\begin{figure*} [!thb]
\begin{center}
\includegraphics[width=5cm]{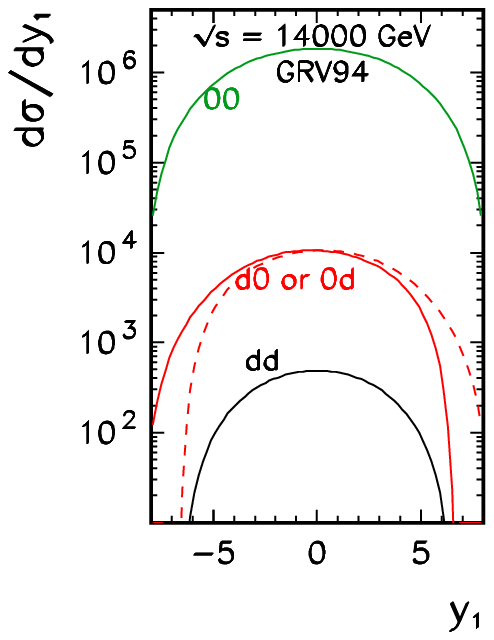}
\includegraphics[width=5cm]{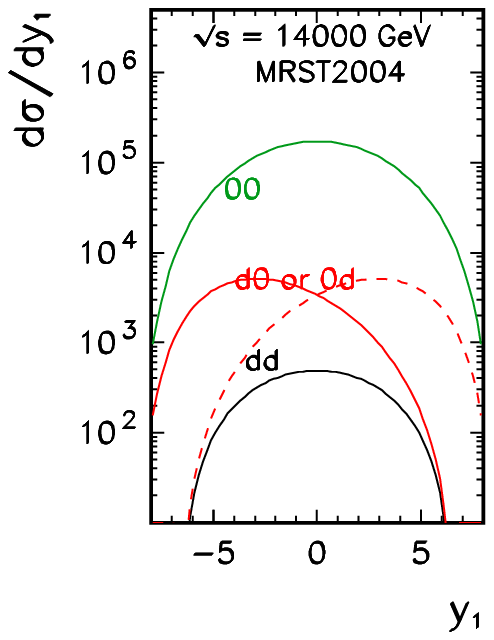}
\includegraphics[width=5cm]{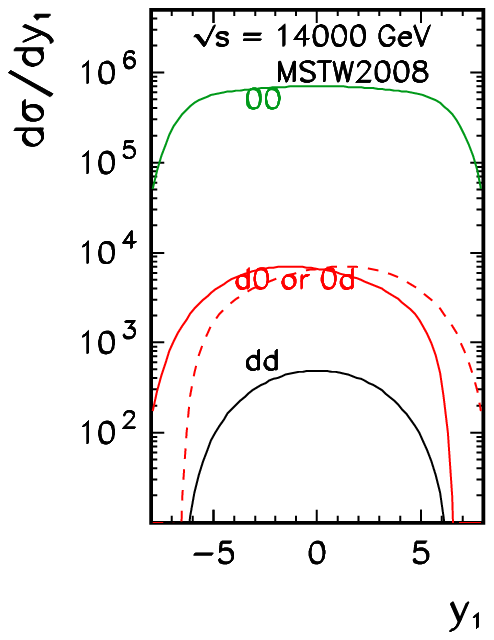}
\end{center}
\caption{\small
Rapidity distribution of $c$ quarks (antiquarks) for LHC energy 
$\sqrt{s} =$ 14 TeV for three different parton distributions. 
The result for single diffractive (0d or d0), central diffractive (dd) 
mechanisms are compared with the standard gluon-gluon fusion contribution (00).
\label{fig:diff_dsig_dy1_14000}
}
\end{figure*}

In Fig.\ref{fig:diff_dsig_dy1_14000} we show distributions in quark
(antiquark) rapidity. We show separately contributions of two different 
single-diffractive components, which give the same distributions in
transverse momentum in Fig.\ref{fig:diff_dsig_dpt_14000}.
When added together they produce a distribution similar in shape to
the standard inclusive case.
Here different parton distributions functions give similar result.
The distributions for different proton gluon distributions are quite different.
This was already observed when discussing photon induced components
in section \ref{sec:exclusive_production}.


Also two dimensional distributions can be interesting as here different 
mechanisms may occupy different parts of the phase space.

\begin{figure*} [!thb]
\begin{center}
\includegraphics[width=6cm]{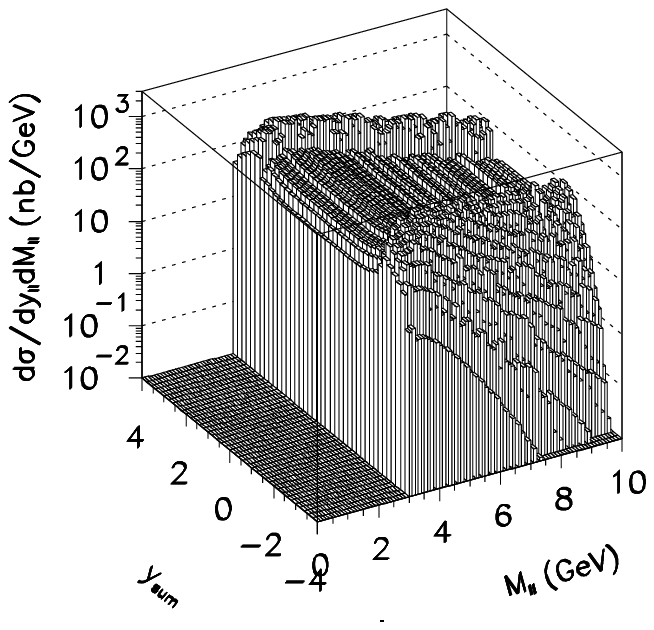}
\includegraphics[width=6cm]{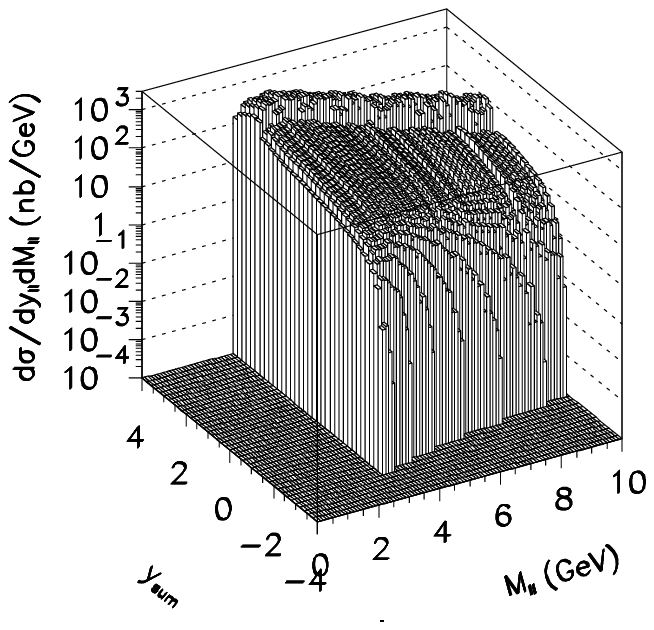}
\includegraphics[width=6cm]{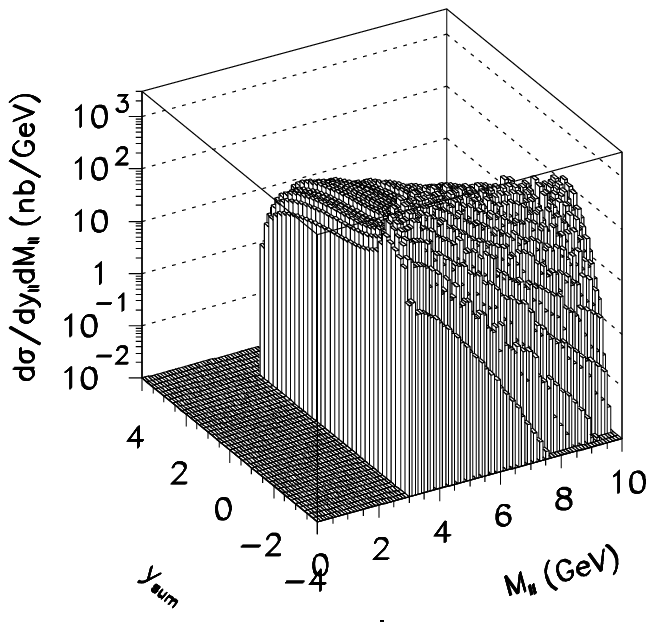}
\includegraphics[width=6cm]{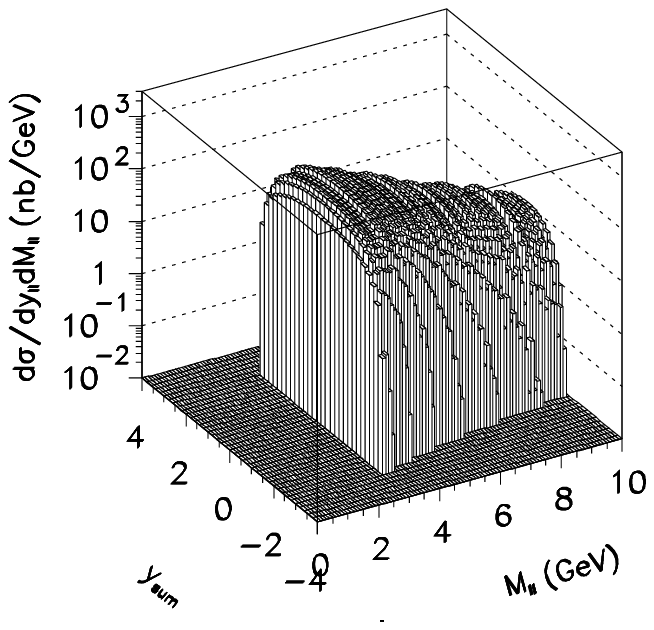}
\end{center}
\caption{\small Two-dimensional distributions in rapidity of the pair
and the quark-antiquark invariant mass for standard (upper left),
single diffractive (upper right and lower left) and central diffractive
contributions. In this calculation MRST04 distributions were used.
\label{fig:YsumMll}
}
\end{figure*}

In Fig.\ref{fig:YsumMll} we show distributions in the rapidity of the
pair and quark-antiquark invariant mass.
Although the distributions are somewhat different the differences occur
in the regions which may be difficult to measure. 
The spread in the pair rapidity for the central diffractive component
is much smaller then that for the inclusive case.


Finally we show distributions in quark and antiquark rapidities.
The distribution for the inclusive central diffractive mechanism are 
concentrated at midrapidities. This is a rather universal feature of 
diffractive processes.


\begin{figure*} [!thb]
\begin{center}
\includegraphics[width=6cm]{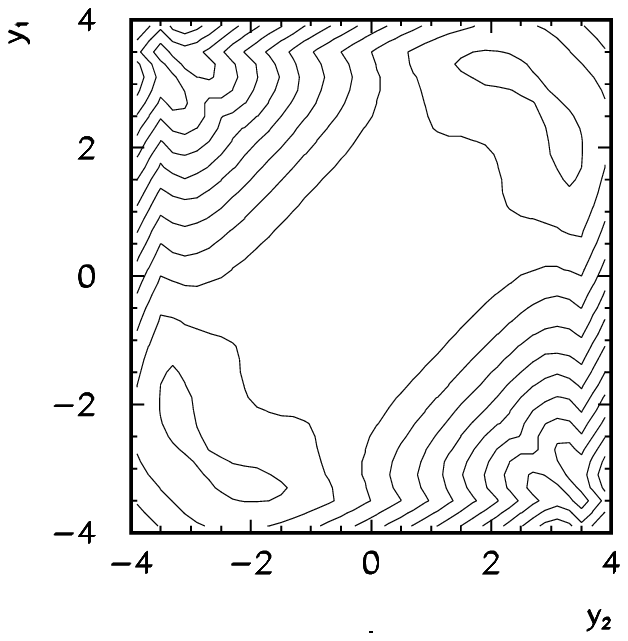}
\includegraphics[width=6cm]{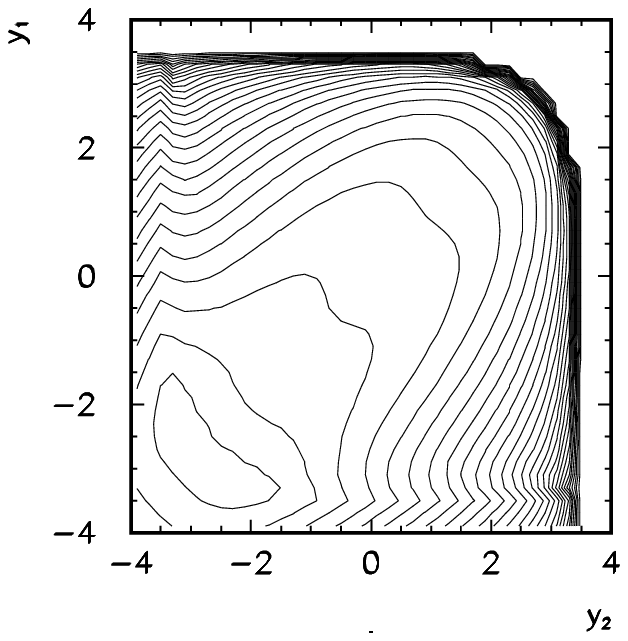}
\includegraphics[width=6cm]{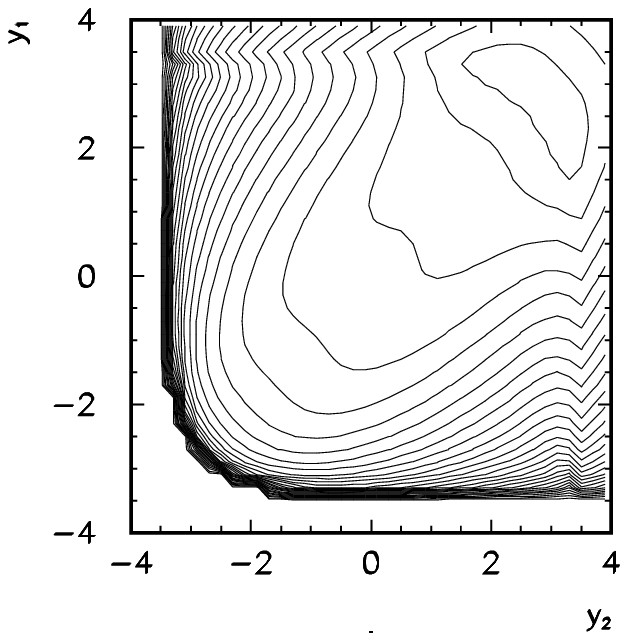}
\includegraphics[width=6cm]{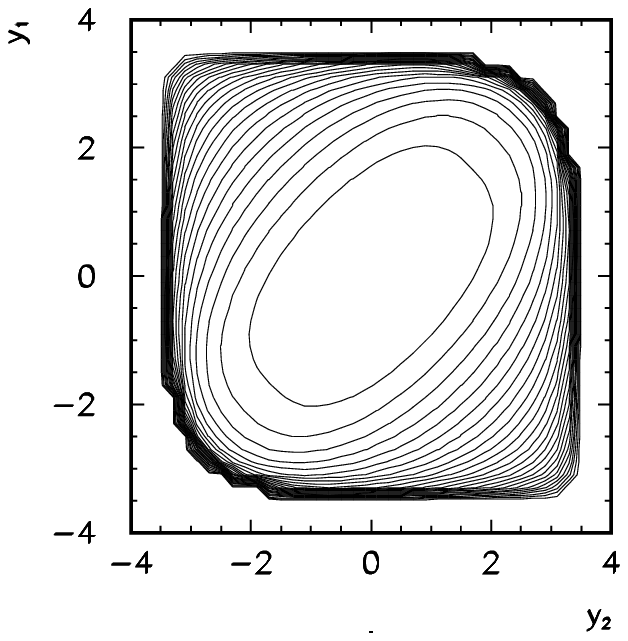}
\end{center}
\caption{\small Two dimensional distribution in rapidity of the quark
and rapidity of the antiquark for standard (upper left),
single diffractive (upper right and lower left) and central diffractive
contributions. In this calculation MRST04 distributions were used.
\label{fig:y1y2}
}
\end{figure*}


The cross section for single and central diffraction is rather small.
However, a very specific final state should allow for its identification
by imposing special conditions on the one-side (single-diffractive
process) or on both-side (central diffractive process) rapidity gaps.
We hope that such an analysis is possible at LHC. Special care
should be devoted to the observation of the exclusive $c \bar c$
production where the observation of $D$ mesons associated by a few pions
would be a proper signal \cite{MS2011}.
Without a special analysis of the final state multiplicity
the exclusive $c \bar c$ production may look like an inclusive
central diffraction. At present there is no analysis of
the final state production for the exclusive $c \bar c$.

A comparison of cross sections for both components will be done in the
next section.

\section{Exclusive central diffractive production of $c \bar c$}

\label{sec:exclusive_production}

There is recently a growing theoretical interest in studying central exclusive
mechanisms of different particles production at high energies, which 
constitute a special category of double-diffractive processes. To date, only a 
few exclusive processes have been measured so far at the Tevatron
collider (see \cite{Albrow} and references therein). 
In particular, central exclusive production (CEP) of the Higgs boson is 
a flag process of special interest and importance in the upcoming Higgs 
searches at the LHC (see e.g. Ref.~\cite{our-bb,our-higgs}).

Generally, in the case of the central exclusive production 
$pp \rightarrow p X p$ with the leading protons, the central system $X$ 
should necessarily be produced in the color singlet state,
such that the proton remnants and the $X$ system are disconnected in 
the color space and their hadronisation occurs independently giving rise
to rapidity gaps \cite{GI}. From the experimental point of view,
CEP procesess are very attractive, because of the rare clean
experimental environment, related to $J_z=0$ selection rule, and great 
mass resolution of the centraly produced object. Such unique features
give a new possibility to exploit $b \bar b$ high branching ratio decay 
channel of the Higgs boson, which is rather impossible in standard inclusive 
measurements, due to very large QCD background.  
Therefore, the QCD mechanism of central exclusive heavy quark dijets 
is a source of the irreducible background to the exclusive Higgs boson 
production.

\begin{figure}[!h]
\includegraphics[width=6cm]{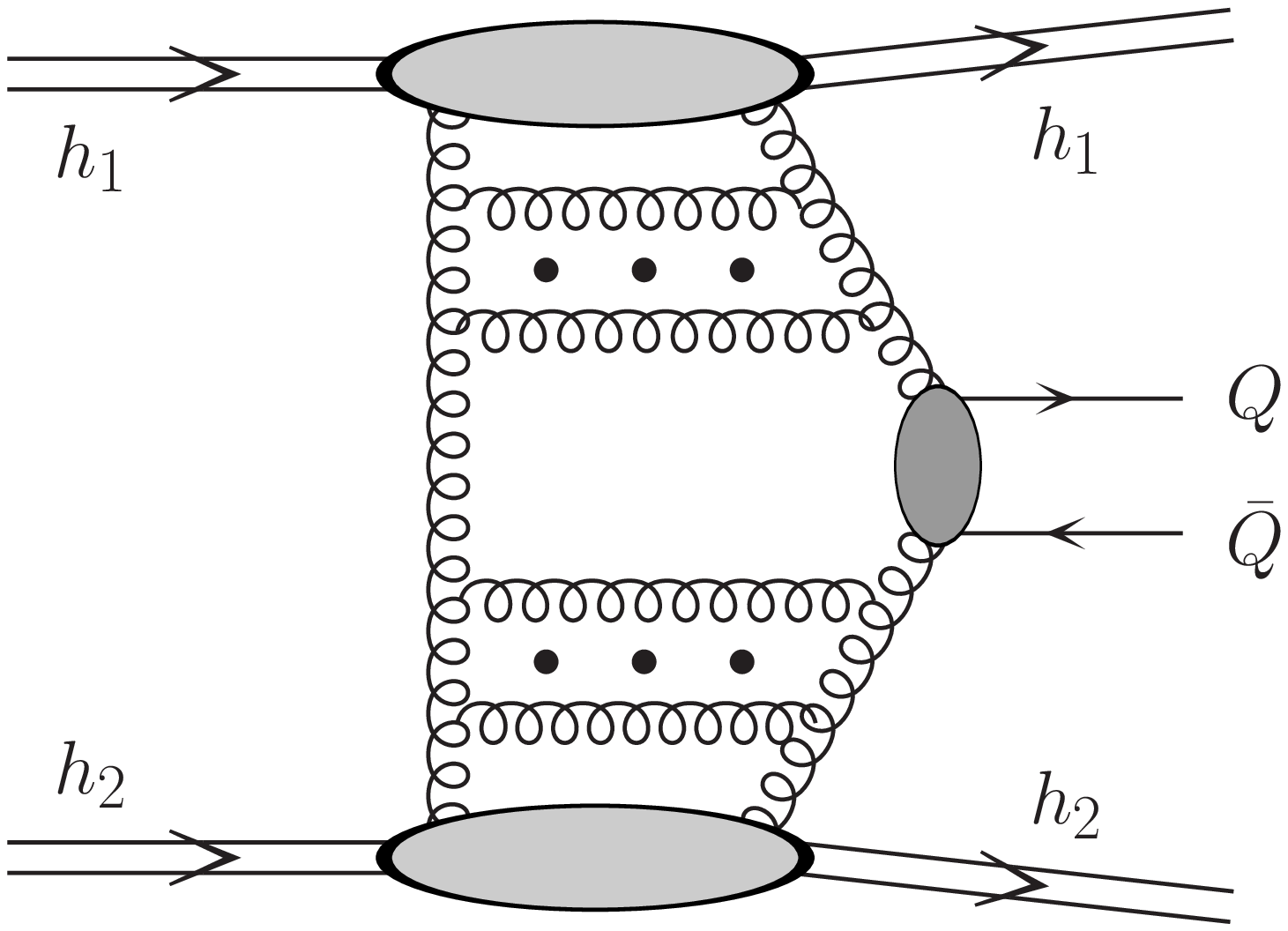}
\includegraphics[width=6cm]{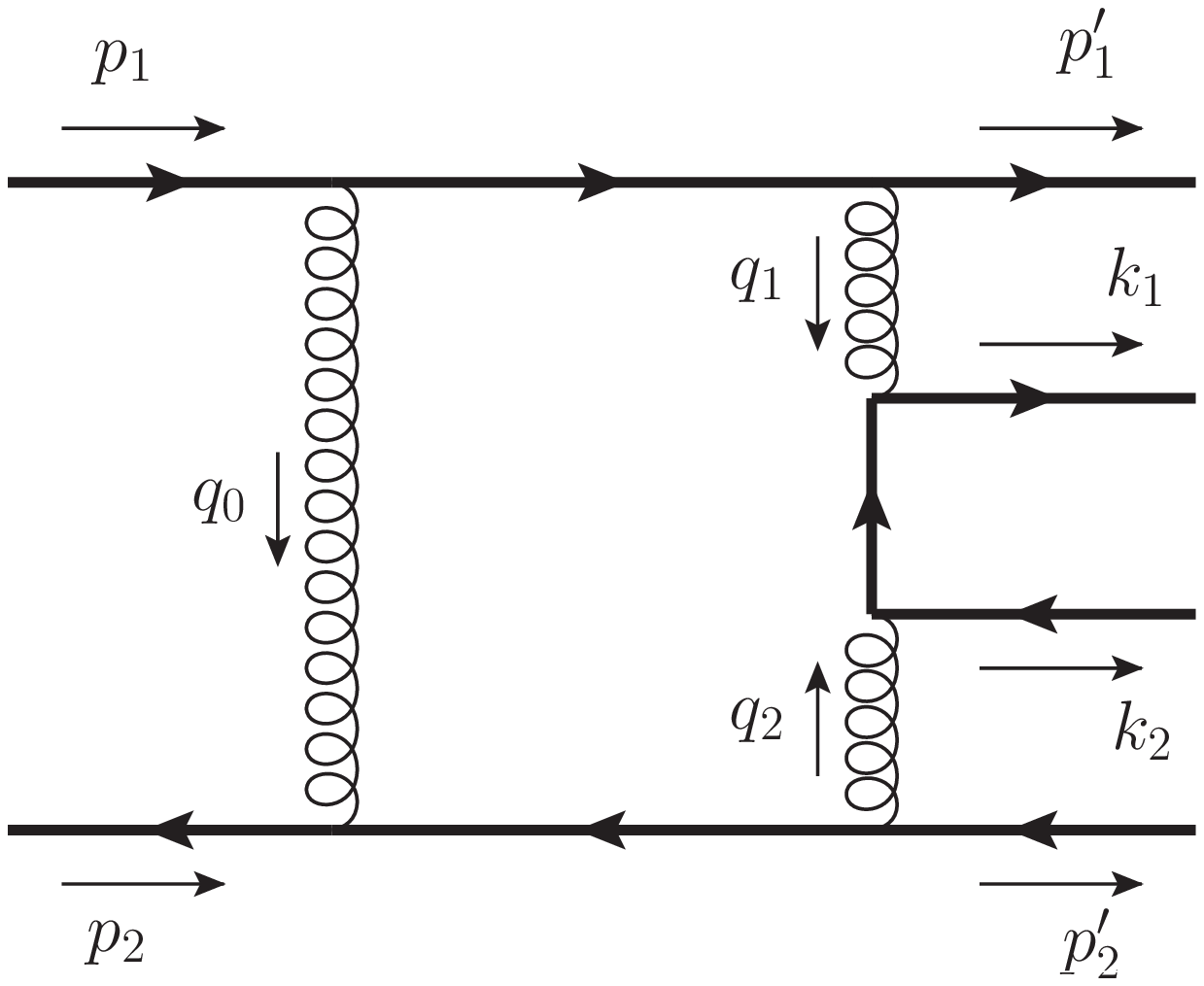}
\caption{\small The mechanism (lef panel) and kinematics (right panel) 
of exclusive double-diffractive production of heavy quarks.
\label{fig:EDD_mechanism}
}
\end{figure}

Central exclusive production of $c \bar c$ and $b \bar b$ pairs was 
studied in detail in our previous papers \cite{MPS10, our-bb,
our-higgs}. In these calculations the $pp\rightarrow p (q \bar q) p$
reaction, illustrated in Fig.~\ref{fig:EDD_mechanism},  was considered 
as a genuine 4-body process with exact kinematics. The applied
perturbative model of theoretical predictions is based on the 
Khoze-Martin-Ryskin (KMR) approach used previously for the exclusive
Higgs boson production \cite{KMR_Higgs}. 
Total cross sections and differential distributions for heavy quarks are
calculated by using $k_{t}$-factorization approach with help of the KMR
unintegrated gluon distribution functions.

This QCD model works very good in the case of exclusive dijets and 
charmonia production, what
was confirmed by CDF data \cite{DKRS2011,MPS_dijets,PST_chic0,PST_chic12}. 
However, 
estimated uncertainties related to gluon densities, factorization and 
renormalization scales, as well as due to absorption corrections are
quite large. It makes the situation somewhat clouded and prevents definite
conlcusions, especially in the case of the exclusive production of heavy
quark pairs. In this context, the promising idea, how to clarify and 
calibrate purely-known parameters of the theoretical model,
is to study $c \bar c$ cross section by exclusive measurements of 
$D \bar D$ meson pairs.
Such experimental studies are being performed now at Tevatron and 
could be also available in Run II experiments at RHIC and at LHC.

Therefore, it is also very interesting, from both, theoretical and 
experimental side, to compare mechanism of central exclusive production
of charm quarks with standard single and double diffractive processes.
Such an analysis of differential cross sections has never been done
before but could bring important informations about differences in 
kinematics and in production rates between them, what is
crucial for future measurements.

According to the KMR approach 
\cite{KMR_Higgs, KMR_Higgs_bbbar_background, KMR-bb}
we write the amplitude of the exclusive diffractive $q\bar{q}$ pair 
production $pp\to p(q\bar{q})p$ as
\begin{eqnarray}
{\cal M}_{\lambda_q\lambda_{\bar{q}}}=
\frac{s}{2}\cdot\frac{\pi^2\delta_{c_1c_2}}{N_c^2-1}\, \Im\int d^2
q_{0,t} \; V_{\lambda_q\lambda_{\bar{q}}}^{c_1c_2}(q_1, q_2) \nonumber 
\times\frac{f^{\mathrm{off}}_{g,1}(x_1,x_1',q_{0,t}^2,
q_{1,t}^2,t_1)f^{\mathrm{off}}_{g,2}(x_2,x_2',q_{0,t}^2,q_{2,t}^2,t_2)}
{q_{0,t}^2\,q_{1,t}^2\, q_{2,t}^2}, \label{amplitude}
\end{eqnarray}
where $\lambda_q,\,\lambda_{\bar{q}}$ are helicities of heavy $q$
and $\bar{q}$, respectively, $t_{1,2}$ are the momentum transfers
along each proton line, $q_{1,t}, q_{2,t}, x_{1,2}$ and
$q_{0,t},\,x_1'\sim x_2'\ll x_{1,2}$ are the transverse momenta and
the longitudinal momentum fractions for active and screening gluons,
respectively. Above $f^{\mathrm{off}}_{g,1/2}$ are the off-diagonal
UGDFs related to both nucleons. The vertex factor
$V_{\lambda_q\lambda_{\bar{q}}}^{c_1c_2}(q_1, q_2) = V_{\lambda_q\lambda_{\bar{q}}}^{c_1c_2}(q_1, q_2, k_1, k_2)$
is the production amplitude of a pair of massive quark $q$ and antiquark $\bar q$ with helicities 
$\lambda_q$, $\lambda_{\bar{q}}$ and momenta $k_1$, $k_2$, respectively.
The longitudinal momentum fractions of active gluons
are calculated based on kinematical variables of outgoing quark
and antiquark:
$x_1 = \frac{m_{q,t}}{\sqrt{s}} \exp(+y_q)
     +  \frac{m_{\bar q,t}}{\sqrt{s}} \exp(+y_{\bar q})$ and
$x_2 = \frac{m_{q,t}}{\sqrt{s}} \exp(-y_q)
     +  \frac{m_{\bar q,t}}{\sqrt{s}} \exp(-y_{\bar q})$,
where $m_{q,t}$ and $m_{\bar q,t}$ are transverse masses of the quark and
antiquark, respectively, and $y_q$ and $y_{\bar q}$ are corresponding
rapidities.

The off-diagonal UGDFs are written as \cite{KMR-ugdf}
 \begin{equation}
 f^{\mathrm{off}}_g(x',x_{1,2},q_{1,2t}^2,q_{0,t}^2,\mu_F^2)\simeq
 R_g\,f_g(x_{1,2},q_{1,2t}^2,\mu_F^2),
 \label{rg}
 \end{equation}
where $R_g\simeq 1.2$ accounts for the single $\log Q^2$
skewed effect \cite{Shuvaev:1999ce}. The factor $R_g$ here cannot be 
calculated from first principles in the most general case of 
off-diagonal UGDFs. It can be estimated only in the case of off-diagonal
collinear PDFs
when $x' \ll x$ and $x g = x^{-\lambda}(1-x)^n$ and then $R_g = \frac{2^{2\lambda+3}}{\sqrt{\pi}}
\frac{\Gamma(\lambda+5/2)}{\Gamma(\lambda+4)}$. In the considered kinematics
the diagonal unintegrated densities can be written in terms of the
conventional (integrated) densities $xg(x,q_t^2)$ as~\cite{KMR-ugdf}
\begin{equation}\label{ugdfkmr}
f_g(x,q_t^2,\mu^2)=\frac{\partial}{\partial\ln q_t^2}
\big[xg(x,q_t^2)\sqrt{T_g(q_t^2,\mu^2)}\big] \; ,
\end{equation}
where $T_g$ is the conventional Sudakov survival factor which
suppresses real emissions from the active gluon during the
evolution, so the rapidity gaps survive.

In the framework of the $k_t$-factorization approach
\cite{ktfac} the hard subprocess $g^*g^*\to q\bar q$ gauge invariant
amplitude $V_{\lambda_q\lambda_{\bar{q}}}^{c_1c_2}(q_1, q_2)$ reads
\begin{eqnarray}\label{qqamp}
&&V_{\lambda_q\lambda_{\bar{q}}}^{c_1c_2}(q_1, q_2)\equiv
n^+_{\mu}n^-_{\nu}V_{\lambda_q\lambda_{\bar{q}}}^{c_1c_2,\,\mu\nu}(q_1, q_2, k_1, k_2),\quad
n_{\mu}^{\mp}=\frac{p_{1,2}^{\mu}}{E_{p,cms}},\\
&&V_{\lambda_q\lambda_{\bar{q}}}^{c_1c_2,\,\mu\nu}(q_1, q_2)=-g_s^2\sum_{i,k}\left\langle
3i,\bar{3}k|1\right\rangle\bar{u}_{\lambda_q}(k_1)\times
(t^{c_1}_{ij}t^{c_2}_{jk}b^{\mu\nu}(k_1,k_2)-
t^{c_2}_{kj}t^{c_1}_{ji}\bar{b}^{\mu\nu}(k_2,k_1))v_{\lambda_{\bar{q}}}(k_2),
\nonumber
\end{eqnarray}
where $E_{p,cms}=\sqrt{s}/2$ is the c.m.s. proton energy, $t^c$ are
the color group generators in the fundamental representation,
$u(k_1)$ and $v(k_2)$ are on-shell quark and antiquark spinors,
respectively, $b^{\mu\nu}$ and $\bar{b}^{\mu\nu}$ are the effective 
vertices arising from the Feynman rules in quasi-multi-Regge kinematics 
(QMRK) approach \cite{FL96}:
\begin{eqnarray} \label{bb}
b^{\mu\nu}(k_1,k_2)=\gamma^{\nu}\frac{\hat{q}_{1}-\hat{k}_{1}-m_q}{(q_1-k_1)^2-m^2}
\gamma^{\mu}-\frac{\gamma_{\beta}\Gamma^{\mu\nu\beta}(q_1,q_2)}{(k_1+k_2)^2}
\; , \\
\bar{b}^{\mu\nu}(k_2,k_1)=\gamma^{\mu}\frac{\hat{q}_{1}-\hat{k}_{2}+m_q}{(q_1-k_2)^2-m^2}
\gamma^{\nu}-\frac{\gamma_{\beta}\Gamma^{\mu\nu\beta}(q_1,q_2)}{(k_1+k_2)^2}
\; , \nonumber
\end{eqnarray}
where $\Gamma^{\mu\nu\beta}(q_1,q_2)$ is the effective three-gluon
vertex. The effective $ggg$-vertices are
canceled out when projecting the $q\bar{q}$ production amplitude
Eq.~(\ref{qqamp}) onto the color singlet state. Since we will adopt the definition of
gluon polarization vectors proportional to transverse momenta
$q_{1/2\perp}$, i.e. $\varepsilon_{1,2}\sim q_{1/2\perp}/x_{1,2}$
(see below), then we must take into account the longitudinal momenta
in the numerators of effective vertices (see Eq.~(\ref{bb})).

The SU(3) Clebsch-Gordan coefficient $\left\langle
3i,\bar{3}k|1\right\rangle=\delta^{ik}/\sqrt{N_c}$ in
Eq.~(\ref{qqamp}) projects out the color quantum numbers of the
$q\bar{q}$ pair onto the color singlet state. Factor $1/\sqrt{N_c}$
provides the averaging of the matrix element squared over
intermediate color states of quarks.

Therefore, we have the following amplitude
\begin{eqnarray}
&&{}V_{\lambda_q\lambda_{\bar{q}}}^{c_1c_2,\,\mu\nu}=-\frac{g_s^2}{2}\,
\delta^{c_1c_2}\,\bar{u}_{\lambda_q}(k_1)
\biggl(\gamma^{\nu}\frac{\hat{q}_{1}-\hat{k}_{1}-m}
{(q_1-k_1)^2-m^2}\gamma^{\mu}-\gamma^{\mu}\frac{\hat{q}_{1}-
\hat{k}_{2}+m}{(q_1-k_2)^2-m^2}\gamma^{\nu}\biggr)v_{\lambda_{\bar{q}}}(k_2)\nonumber.\\
 \label{vector_tensor}
\end{eqnarray}

In the present calculations we use the GJR08 set of collinear gluon
distributions \cite{GJR}. In the analogy to the CEP of Higgs boson,
where renormalization and factorization scales are advocated to be 
$\mu^{2} = \mu_{R}^2= \mu_{F}^2 = M_{H}^{2}$ \cite{CF2010}, we 
apply the following prescription $\mu^{2} = M_{c \bar c}^2$. Absorption 
corrections to the bare $p p \rightarrow p (q \bar q) p$ amplitude,
which are necessary to be taken into account (to ensure exclusivity of 
the process), are included approximately by multiplying the cross section 
by the gap survival factors $S_G = 0.1$ for RHIC and $S_G = 0.03$ for 
the LHC energy.
More details about exclusive production of heavy quarks can be found in
our original paper \cite{MPS10}.
Let us come now to presentation of our results.

\begin{figure} [!thb]
\begin{center}
\includegraphics[width=6cm]{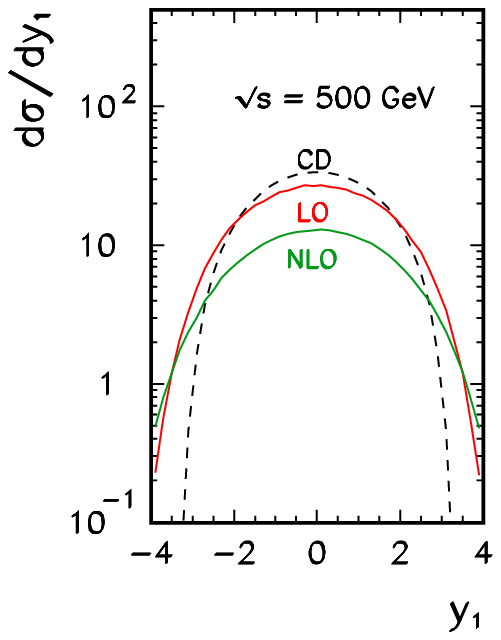}
\includegraphics[width=6cm]{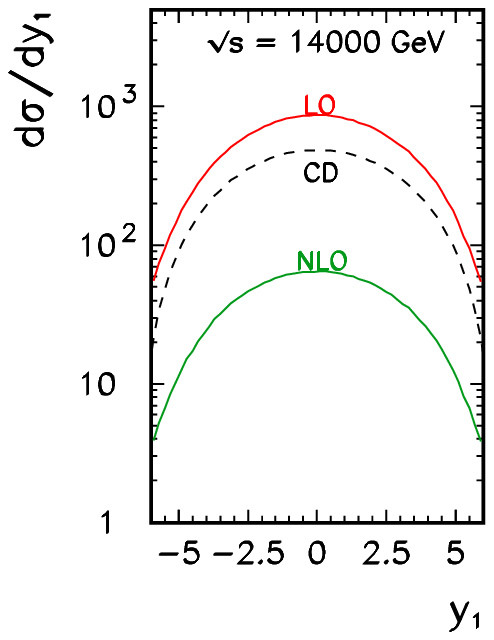}
\end{center}
\caption{\small Distributions in rapidity 
of $c$ quark/antiquark for the exclusive component 
at $\sqrt{s}$ = 500 GeV (left panel)
and $\sqrt{s}$ = 14 TeV (right panel).
For comparison we show the central diffractive contribution (dashed line).
Different collinear gluon distributions were used to obtain
the unintegrated gluon distribution according to the KMR prescription.
\label{fig:dsig_dy_exc}
}
\end{figure}

In Fig.\ref{fig:dsig_dy_exc} we show rapidity distribution
of $c$ quarks from the exclusive mechanism shown in 
Fig.~\ref{fig:EDD_mechanism} (solid line). We show the results
for leading order (upper curves) and next-to-leading order
collinear gluon distributions \cite{GJR}.
We observe large difference of results for LO and NLO gluon distribution
especially at LHC.
For comparison we show the contribution of central diffractive component
discussed in section \ref{sec:diffractive_production}.
In this calculation we have included gap survival factors
$S_G$ = 0.1 for $\sqrt{s}$ = 500 GeV and $S_G$ = 0.03 for $\sqrt{s}$ =
14 TeV.
The cross section for the exclusive mechanism is similar to that 
for the inclusive central diffractive mechanism. 
The exclusive production starts to dominate only at large $c$ quark 
rapidities. Therefore a measurement of the cross
section with double (both side) rapidity gaps may be not sufficient
to single out the exclusive mechanism. Clearly other cuts would be
necessary.

\begin{figure} [!thb]
\begin{center}
\includegraphics[width=6cm]{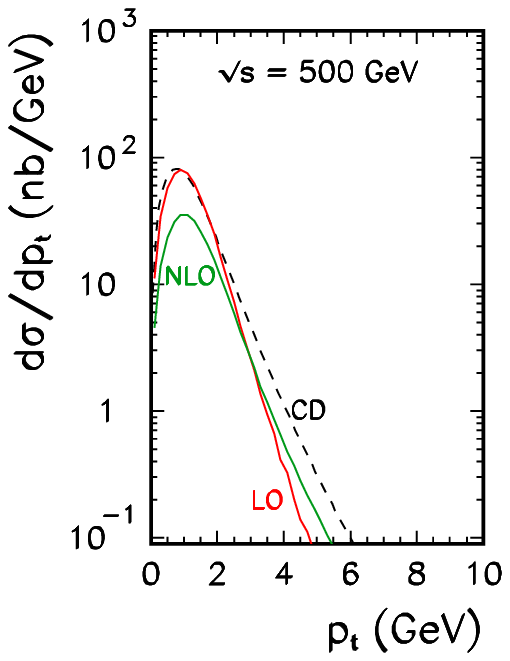}
\includegraphics[width=6cm]{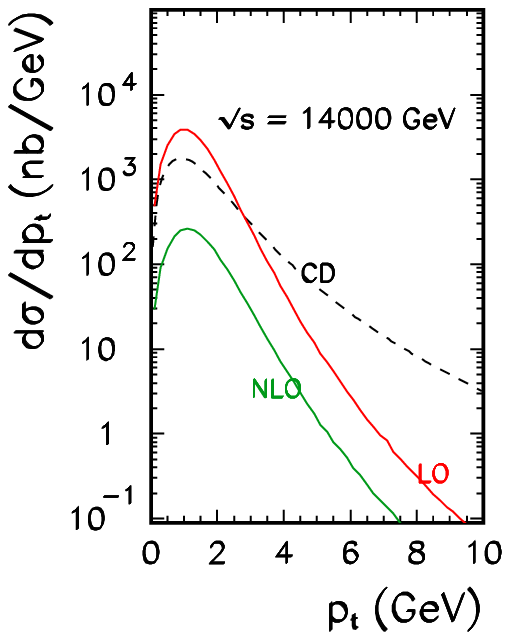}
\end{center}
\caption{\small Distributions in transverse momentum 
of $c$ quark/antiquark for the exclusive component 
at $\sqrt{s}$ = 500 GeV (left panel)
and $\sqrt{s}$ = 14 TeV (right panel) TeV.
Different collinear gluon distributions were used to obtain
the unintegrated gluon distribution according to the KMR prescription.
For comparison we show the inclusive central diffractive contribution 
(dashed line).
\label{fig:dsig_dpt_exc}
}
\end{figure}

Corresponding distributions in the $c$ quark ($ \bar c$ antiquark) 
transverse momentum are shown in Fig.\ref{fig:dsig_dpt_exc}.
The distribution for exclusive component extends to higher transverse
momentum than that for the central inclusive diffractive one. 
A lower cut on $c$ quark (D meson) transverse momentum may therefore 
help to identify the exclusive component but will exclude a measurment
of the integrated cross section for this component.
A detailed Monte Carlo studies of final states of both components may 
help to find a better criterion to separate experimentally the two components.

\section{Conclusions}

In the present paper we have calculated differential distributions 
for different subdominant contributions usually neglected in 
the literature when calculating production of $c \bar c$ pairs.

Single and double photon induced processes are first class of mechanisms
considered here. In calculating single particle distributions we have
used a special set of parton distributions which includes photon as 
a parton of the proton. The calculation of the cross section
is therefore very similar to that for the gluon-gluon fusion.
The difference is only in color factors and a lack of the s-channel 
diagrams for photon induced processes.
We have also included contributions when emitted photon which enters
a hard process leaves a proton in a ground state. Those ``elastic'' 
mechanisms give similar contribution as the ``deeply inelastic'' 
mechanisms considered in the QCD-improved parton model. 
We have found that although individual contributions are very small,
when added together they can give cross section of about 1 \%
of the inclusive one dominated by gluon-gluon fusion.
In our analysis we have neglected resonance contributions when 
the photon leaves the remaining object in a proton excited state, 
e.g. in $\Delta$(1220) resonance or other nucleon resonances.

We have also discussed single and central diffractive production
of $c \bar c$ pairs in the Ingelman-Schlein model. In these
calculations we have included diffractive parton distributions
obtained by the H1 collaboration at HERA and absorption effects
neglected in some early calculations in the literature.
The absorption effects which are responsible for the 
naive Regge factorization breaking cause that the cross section 
for diffractive processes is much smaller than that for the fully 
inclusive case, but could be measured at RHIC and LHC by 
imposing special condition on rapidity gaps.

Finally we have discussed a fully exclusive diffractive production
of $c \bar c$. It was advocated recently that the cross section
for this mechanism may be substantial. We have found here that 
both at RHIC and LHC its contribution is smaller than that for single 
diffractive one.
In our opinion it is very timely to analyze if this contribution 
could be measured. This equires
an analysis of the final state. We expect that the final state
in single and exclusive production are different enough to set
criteria to pin down the fully exclusive component.
It is, however, not obvious if the central diffractive and
purely exclusive mechanisms could be differentiated experimentally. 
They may look similar as far as rapidity gap structure is considered.
We predict that the total contribution
of central diffractive mechanism is similar to that for the exclusive one. 
In contrast the final state multiplicity can be expected to be
different. A better analysis requires a Monte Carlo studies.

We have not discussed an impact of diffractive mechanisms
considered in the present paper on the fully inclusive cross section
for $c \bar c$ pair production. 
This is a rather difficult task and goes beyond the scope of 
the present paper.

\vskip 2cm

{\bf Acknowledgments}

We are indebted to Wolfgang Sch\"afer and Roman Pasechnik for interesting
conversions. This work was partially supported by the polish
grant N N202 237040.

\vskip 1cm


\end{document}